\documentclass[letterpaper,twocolumn,10pt]{article}
\usepackage{usenix}
\usepackage{epsfig}
\usepackage{comment}
\usepackage[pdftex]{color}
\usepackage{url}
\usepackage{listings}
\usepackage{textcomp}
\usepackage{gnuplot-lua-tikz}
\usepackage{amssymb}
\usepackage{bm}
\usepackage{paralist}
\usepackage{ulem}
\usepackage{subcaption}
\definecolor{lbcolor}{rgb}{0.9,0.9,0.9}

\newlength{\saveparindent}
\newlength{\saveparskip}
\begin{document}

%don't want date printed
\date{}

%make title bold and 14 pt font (Latex default is non-bold, 16 pt)
\title{FlashGraph: Processing Billion-Node Graphs %with Multi-Core Parallelism 
on an Array of Commodity SSDs}

%for single author (just remove % characters)
\author{
{\rm Da Zheng, Disa Mhembere, Randal Burns}\\
Department of Computer Science \\
Johns Hopkins University
\and
{\rm Joshua Vogelstein}\\
Institute for Computational Medicine \\
Department of Biomedical Engineering \\
Johns Hopkins University
% copy the following lines to add more authors
\and
{\rm Carey E. Priebe}\\
Department of Applied Mathematics and Statistics \\
Johns Hopkins University
\and
{\rm Alexander S. Szalay}\\
Department of Physics and Astronomy \\
Johns Hopkins University
} % end author

\maketitle

% Use the following at camera-ready time to suppress page numbers.
% Comment it out when you first submit the paper for review.
\thispagestyle{empty}

\subsection*{Abstract}
Graph analysis performs many random reads and writes, thus, these workloads are
typically performed in memory. Traditionally, analyzing large graphs requires
a cluster of machines so the aggregate memory exceeds the graph size.
We demonstrate
that a multicore server can process graphs with billions of vertices and hundreds
of billions of edges, utilizing commodity SSDs with minimal performance loss.
We do so by implementing a graph-processing engine on top of a user-space SSD
file system designed for high IOPS and extreme parallelism. 
Our semi-external memory graph engine called FlashGraph stores vertex state
in memory and edge lists on SSDs. It hides latency by overlapping computation
with I/O. To save I/O bandwidth, FlashGraph only accesses edge lists requested
by applications from SSDs; to increase I/O throughput and reduce CPU overhead for I/O,
it conservatively merges I/O requests. These designs maximize performance
for applications with different I/O characteristics. FlashGraph exposes
a general and flexible vertex-centric programming interface that can express
a wide variety of graph algorithms and their optimizations. We demonstrate that
FlashGraph in semi-external memory performs many algorithms with performance
up to 80\% of its in-memory implementation and significantly outperforms
PowerGraph, a popular distributed in-memory graph engine.

\section{Introduction}
% What problem are you going to solve.
Large-scale graph analysis has emerged as a fundamental computing
pattern in both academia and industry.  This has resulted in
specialized software ecosystems for scalable graph computing in the cloud 
with applications to web structure and social networking \cite{giraph,pregel},
machine learning \cite{graphlab}, and network analysis \cite{Pearce10}.
The graphs are massive: Facebook's social graph has billions of vertices
and today's web graphs are much larger.  

% Why is it hard?
The workloads from graph analysis present great challenges to system designers.
Algorithms that perform edge traversals on graphs induce many small, random
I/Os, because edges encode non-local structure among vertices and many
real-world graphs exhibit a power-law distribution on the degree 
of vertices. As a result, graphs cannot be clustered or partitioned
effectively \cite{Leskovec} to localize access.
While good partitions may be important for performance~\cite{surfer}, 
leading systems partition natural graphs randomly \cite{powergraph}.

%LESKOVEC, J., LANG, K. J., DASGUPTA, A., , AND MAHONEY,
%%M. W. Community structure in large networks: Natural cluster
%sizes and the absence of large well-deﬁned clusters. Internet
%Mathematics 6, 1 (2008), 29–123.

% surfer graph paritioningo http://research.microsoft.com/pubs/131014/surfer_tr.pdf

%How have others addressed the problem?
Graph processing engines have converged on a design that \textit{(i)} stores
graph partitions in the aggregate memory of a cluster, \textit{(ii)} encodes
algorithms as parallel programs against the vertices of the graph, and
\textit{(iii)} uses either distributed shared memory \cite{graphlab,powergraph}
or message passing \cite{pregel,giraph,trinity} to communicate between
non-local vertices.
Placing data in memory reduces access latency when compared to disk drives.
Network performance, required for communication between graph partitions,
emerges as the bottleneck and graph engines require fast networks to realize
good performance.

Recent work has turned back to processing graphs from disk drives on a single
machine \cite{graphchi,xstream} to achieve scalability without excessive hardware.
These engines are optimized for the sequential performance of magnetic disk
drives; they eliminate random I/O by scanning the entire graph dataset.
This strategy can be wasteful for algorithms that access only small fractions
of data during each iteration.  For example, breadth-first search,
a building block
for many graph applications, only processes vertices in a frontier. 
PageRank \cite{pagerank} starts processing all vertices in a graph, but as the 
algorithm progresses, it narrows to a small subset of active vertices.
There is a huge performance gap between these systems and in-memory processing.

%What is the nature of your solution?
%We present a graph-processing engine, called FlashGraph, that meets
We present FlashGraph, a semi-external memory graph-processing engine
that meets or exceeds the performance of in-memory engines and allows graph
problems to scale to the capacity of semi-external memory. Semi-external memory
\cite{Abello98, Pearce10}
maintains algorithmic vertex state in RAM and edge lists
on storage. The semi-external memory model avoids
writing data to SSDs. Only using memory for vertices increases the scalability
of graph engines in proportion to the ratio of edges to vertices in a graph,
more than 35 times for our largest graph of Web page crawls. FlashGraph uses
an array of solid-state
drives (SSDs) to achieve high throughput and low latency to storage.
Unlike magnetic disk-based engines, FlashGraph supports selective
access to edge lists. %Although this paper explores the scalability limits
%of graph processing on a shared-memory multiprocessor, we envision deploying
%this architecture as the building block for graph-processing clusters with the 
%benefit that graph partitions that are orders of magnitude larger will better
%utilize hardware and result in less network traffic.

% Why is it new/different/special?
Although SSDs can deliver high IOPS, we overcome many technical challenges
to construct a semi-external memory graph engine with performance comparable
to an in-memory graph engine. The throughput of SSDs are an order of magnitude
less than DRAM and the I/O latency is multiple orders of magnitude slower.
Also, I/O performance is extremely non-uniform and needs to be localized.
%Processors have affinity to specific I/O controllers
%and I/O controllers to SSDs.  Not respecting these differences can reduce performance by a factor of ten.
%\dz{where does the number of 10 come from?}
Finally, high-speed I/O consumes many CPU cycles, interfering
with graph processing.

We build FlashGraph on top of a user-space SSD file system called SAFS \cite{safs}
to overcome these technical challenges. The set-associative file system
(SAFS) refactors I/O scheduling, data placement, and data caching
for the extreme parallelism of modern NUMA multiprocessors. The lightweight
SAFS cache enables FlashGraph to adapt to graph applications with different
cache hit rates. We
integrate FlashGraph with the asynchronous user-task I/O interface of SAFS to
reduce the overhead of accessing data in the page cache and memory consumption,
as well as overlapping computation with I/O.

FlashGraph issues I/O requests carefully to maximize the performance
of graph algorithms with different I/O characteristics.
It reduces I/O by only accessing edge lists requested by applications and using
compact external-memory data structures. It reschedules I/O access
on SSDs to increase the cache hits in the SAFS page cache.
It conservatively merges I/O requests to increase I/O throughput and reduces
CPU overhead by I/O.

% What are it's key features?
Our results show that FlashGraph in semi-external memory achieves performance
comparable to its in-memory version and Galois \cite{galois}, a high-performance,
in-memory graph engine with a low-level API, on a wide-variety
of algorithms that generate diverse access patterns.
FlashGraph in semi-external memory mode significantly outperforms PowerGraph,
a popular distributed in-memory graph engine.
We further demonstrate that FlashGraph can process massive natural graphs
in a single machine with relatively small memory footprint;
e.g., we perform breadth-first search on a graph of 3.4 billion vertices and
129 billion edges using only 22 GB of memory.
Given the fast performance and small memory footprint, we conclude that
FlashGraph offers unprecedented opportunities for users to perform massive graph
analysis efficiently with commodity hardware.

\section{Related Work}
MapReduce \cite{mapreduce} is a general large-scale data processing framework.
PEGASUS \cite{pegasus} is a popular graph processing engine whose
architecture is built on MapReduce.
PEGASUS respects the nature of the MapReduce programming paradigm and expresses
graph algorithms as a generalized form of sparse matrix-vector multiplication.
This form of computation works relatively well for graph algorithms such as
PageRank \cite{pagerank} and label propagation \cite{label_prop}, but performs
poorly for graph traversal algorithms.

%There are many other works \cite{linear_algebra, kdt} that perform graph analysis
%Other frameworks \cite{linear_algebra, kdt} perform graph analysis
%in the form of linear algebra. They represent a graph in a sparse
%matrix and vertex states in a vector. In this abstraction, they express
%PageRank and label propagation
%with sparse matrix dense vector multiplication and breadth-first search as sparse
%matrix sparse vector multiplication.  Our work follows the alternative
%design strategy that expresses graph algorithms as parallel program
%against the vertices of the graph, which suits graph traversal algorithms better.

Several other works \cite{linear_algebra, kdt} perform graph analysis
using linear algebra with sparse adjacency matrices and vertex-state 
vectors as data representations.
In this abstraction, PageRank and label propagation are efficiently expressed
as sparse-matrix, dense-vector multiplication, and breadth-first search as 
sparse-matrix, sparse-vector multiplication. These frameworks target
mathematicians and those with the ability to formulate and express their
problems in the form of linear algebra.

%Boost Graph Library \cite{bgl}, Parallel Boost Graph Library \cite{pbgl} and
%iGraph \cite{igraph} provide a large collection of pre-written graph algorithms.
%Despite users benefiting from these libraries by invoking existing implementations
%and using the APIs, these libraries lack a runtime environment within which
%to express natively parallel algorithms that scale to match very large-scale graphs.

%Boost Graph Library \cite{bgl}, Parallel Boost Graph Library \cite{pbgl} and
%iGraph \cite{igraph} provide a large collection of existing graph algorithms.
%\rb{This feels to negative.  We shouldn't call them limited.  We should instead
%find a way to express these as a different class of solutions that are not comparable 
%to what we are doing.}
%Users can benefit from these graph libraries by invoking the existing
%implementations of graph algorithms, but they provide limited programming API
%for users to express their own graph algorithms.

Pregel \cite{pregel} is a distributed graph-processing framework that 
%graph analysis. It 
allows users to express graph algorithms in vertex-centric programs  
using bulk-synchronous processing (BSP).
It abstracts away the complexity of programming in a distributed-memory 
environment and runs users' code in parallel on a cluster.
Giraph \cite{giraph} is an open-source implementation of Pregel.

Many distributed graph engines adopt the vertex-centric programming model
and express different designs to improve performance.
GraphLab \cite{graphlab} and PowerGraph \cite{powergraph} prefer shared-memory
to message passing and provide asynchronous execution. FlashGraph supports
both pulling data from SSDs and pushing data with message passing.
FlashGraph does provide asynchronous
execution of vertex programs to overlap computing with data access. 
Trinity \cite{trinity} optimizes message passing by restricting vertex
communication to a vertex and its direct neighbors.

Ligra \cite{ligra} is a shared-memory graph processing framework and
its programming interface is specifically optimized for graph traversal algorithms.
It is not as general as other graph engines
such as Pregel, GraphLab, PowerGraph, and FlashGraph. Furthermore, Ligra's 
maximum supported graph size is limited by the memory size of a single machine.

Galois \cite{galois} is a graph programming framework with a low-level abstraction
to implement graph engines. The core of the Galois framework is its novel task
scheduler. The dynamic task scheduling in Galois is orthogonal to FlashGraph's
I/O optimizations and could be adopted.  
% In contrast, our scheduler in FlashGraph
%mainly focuses on optimizing I/O: increase the cache hit rate and the I/O
%request size.

GraphChi \cite{graphchi}
and X-stream \cite{xstream} are specifically designed for magnetic disks. They
eliminate random data access from disks by scanning the entire graph dataset
in each iteration. Like graph processing frameworks built on top of MapReduce,
they work relatively well for graph algorithms that require computation on all
vertices, but share the same limitations, i.e., suboptimal graph traversal
algorithm performance.

TurboGraph \cite{turbograph} is an external-memory graph
engine optimized for SSDs. Like FlashGraph, it reads vertices selectively
and fully overlaps I/O and computation. TurboGraph targets graph algorithms
expressed in sparse matrix vector multiplication, so it is difficult to
implement graph applications such as triangle counting. It uses much larger
I/O requests than FlashGraph to read vertices selectively due to its
external-memory data
representation. Furthermore, it targets graph analysis on a single SSD or
a small SSD array and does not aim at performance comparable to
in-memory graph engines.

Abello et al. \cite{Abello98} introduced the semi-external memory algorithmic
framework for graphs.  Pearce et al.~\cite{Pearce10} implemented several 
semi-external memory graph traversal algorithms for SSDs.
FlashGraph adopts and advances several concepts introduced by these works.

%FlashGraph uses a different programming model from GraphLab and PowerGraph.
%Although GraphLab and PowerGraph work in the distributed memory setting, they
%expose shared-memory data structures to programmers. In contrast, FlashGraph,
%although it works in the shared memory setting, uses a share-nothing model. 
%FlashGraph uses a programming model similar to Pregel but with some improvements
%for out-of-core graph analysis.

%Existing frameworks such as GraphLab, GraphChi employ a vertex-centric
%Gather-Apply-Scatter programming model. Despite working in the 
%distributed memory setting, these packages expose shared-memory data structures
%to programmers causing degradation in performance as synchronization is 
%frequently necessary to avoid race conditions. 

%In contrast, FlashGraph,uses a share-nothing model while additionally possessing
%the ability to work in the shared memory setting as well. Hence, FlashGraph 
%uses a programming model similar to Pregel, but with several critical improvements
%for out-of-core graph analysis. For example, we only maintain the 
%user-defined vertex state in memory, leaving edges on disk, thus substantially reducing
%memory consumption. We instead use message passing for communication between vertices.

\section{Design}
%FlashGraph is a semi-external memory graph engine optimized for high-speed
%flash drives. It stores the neighbor list of vertices in the flash devices
FlashGraph is a semi-external memory graph engine optimized for 
any fast I/O device such as Fusion I/O or arrays of solid-state drives
(SSDs). It stores the edge lists of vertices on SSDs and maintains
vertex state in memory. FlashGraph runs on top of the set-associative
file system (SAFS) \cite{safs}, a user-space filesystem designed to realize
both high IOPS and lightweight caching for SSD arrays on non-uniform memory
and I/O systems.

%FlashGraph utilizes SAFS to perform high-throughput parallel asynchronous 
%I/O from SSDs and  page caching.

We design FlashGraph with two goals: to achieve performance comparable to
in-memory graph engines while realizing the increased scalability of
the semi-external memory execution model; to have a concise and flexible
programming interface to express
a wide variety of graph algorithms, as well as their optimizations.
%\item to minimize the amortized overhead of constructing a graph for
%analysis in FlashGraph.

To optimize performance, we design FlashGraph with the following principles:
	
%	\vspace{2 mm}
	\noindent \textbf{Reduce I/O}:
	Because SSDs are an order of magnitude slower than RAM, FlashGraph saturates
	the I/O channel in many graph applications. Reducing the amount of I/O for
	a given algorithm directly
	improves performance.  FlashGraph \textit{(i)} compacts data structures,
	\textit{(ii)} maximizes cache hit rates and \textit{(iii)} performs selective
	data access to edge lists.

%  \vspace{2 mm}
	\noindent \textbf{Perform sequential I/O when possible}: Even though SSDs
	provide high IOPS for random access, sequential I/O always outperforms
	random I/O and reduces the CPU overhead of I/O processing in the kernel.

%  \vspace{2 mm}
%	\noindent \textbf{Maximize cache hit rates}: 
%    The high-speed I/O of SSDs is still an order of magnitude slower than RAM. 
%    I/O will be the bottleneck if SSDs serve all data for graph processing.
%    Careful scheduling by the graph engine orders data accesses to increase
%    data reuse in the page cache.

%  \vspace{2 mm}
%	\noindent \textbf{Reduce random memory access:} Random access in RAM reduces the 
%  effectiveness of CPU caches and decreases memory bandwidth. 
%  It is as important to access vertices
%  sequentially (from memory) as it is to access edges sequentially (from SSDs).

%  \vspace{2 mm}
	\noindent \textbf{Overlap I/O and Computation}: To fully utilize multicore processors
   and SSDs for data-intensive workloads, one must initiate many parallel I/Os and 
   process data when it is ready. 
%  \item {\em Avoid remote memory access:} Modern multi-proessor systems 
%  have non-uniform memory architectures (NUMA) in which regions of memory 
%  associate with processors.  Accessing remote memory (of another processor) 
%  has higher latency, lower bandwidth, and causes overhead and 
%  contention on the remote processor.  

%  \vspace{2 mm}
   \noindent \textbf{Minimize wearout}: SSDs wear out after many writes,
		especially for consumer SSDs.
	Therefore, it is important to minimize writes to SSDs. This includes avoiding
	writing data to SSDs during the application execution and reducing the necessity
	of loading graph data to SSDs multiple times for the same graph.

In practice, selective data access and performing sequential I/O
conflict. Selective data access prevents us from generating large sequential
I/O, while using large sequential I/O may bring in unnecessary data from SSDs
in many graph applications. For SSDs, FlashGraph places a higher priority in
reducing the number of bytes read from SSDs than in performing sequential I/O
because the random (4KB) I/O throughput
of SSDs today is only two or three times less than their sequential I/O.
In contrast, hard drives have random I/O throughput two orders of
magnitude smaller than their sequential I/O. Therefore, other external-memory graph
engines such as GraphChi and X-stream place a higher priority in performing
large sequential I/O.

%FlashGraph realizes these design principles through a close integration
%with the SAFS file system.  It pushes data-oriented computing into the
%SAFS page cache, which allows it to use asynchronous I/O interfaces
%for parallel I/O and pushes processing to the data location, removing
%copies and data movement.

\subsection{SAFS}
% Missing a topic sentence here. IMO necessary. Revise as necessary
SAFS \cite{safs} is a user-space filesystem for high-speed SSD arrays in
a NUMA machine. It is implemented as a library and runs in the address
space of its application. It is deployed on top of the Linux native filesystem.

SAFS reduces overhead in the Linux block subsystem, enabling maximal
performance from an SSD array. It deploys dedicated per-SSD I/O threads to
issue I/O requests with Linux AIO to reduce locking overhead in the Linux kernel;
it refactors I/Os from applications and sends them to I/O threads
with message passing. Furthermore, it has
a scalable, lightweight page cache that organizes pages in a hashtable and
places multiple pages in a hashtable slot \cite{SA-cache}. This page cache
reduces locking overhead and
incurs little overhead when the cache hit rate is low; it increases
application-perceived performance linearly along with the cache hit rate.

%SAFS is optimized for both small and large I/O requests. Natural graphs
%exhibit a power-law distribution on vertex degrees, so graph analysis
%requires the access of a few large vertices and many small vertices.
%Therefore, SAFS preserves large I/O requests from the upper layer
%and merges small requests into larger ones when they go through SAFS's page
%cache.
%Large requests increase I/O throughput and reduce CPU overhead.

To better support FlashGraph, we add an asynchronous user-task I/O interface
to SAFS. This I/O interface supports general-purpose computation in the page cache, avoiding
the pitfalls of Linux asynchronous I/O.  
To achieve maximal performance, SSDs require many parallel I/O requests.
This could be achieved with user-initiated asynchronous I/O.  However, 
this asynchronous I/O requires the allocation of user-space buffers in advance
and the copying of data into these buffers. This creates processing overhead
from copying and further pollutes memory with empty buffers waiting to be filled.
When an application issues a large number of parallel I/O requests,
the empty buffers account for substantial memory consumption.
In the SAFS user-task programming interface, an application associates
a user-defined task with each I/O request.
Upon completion of a request, the associated user task executes
inside the filesystem, accessing data in the page
cache directly. Therefore, there is no memory allocation and copy
for asynchronous I/O.

%The SAFS user-task interface provides a Java-style iterator that abstracts out
%the details of non-contiguous data access. In the user-task I/O interface,
%functions operate on multiple pages
%of data in the cache without copying them into a buffer.
%However, accessing data in discontiguous pages requires extra CPU instructions.
%Our Java-style
%iterator allows functions to read the data across multiple, discontiguous pages
%as if they were sequential, while minimizing the number of CPU instructions used
%for data access.

\subsection{The architecture of FlashGraph}
\begin{figure}[t]
\centering
\includegraphics[scale=0.3]{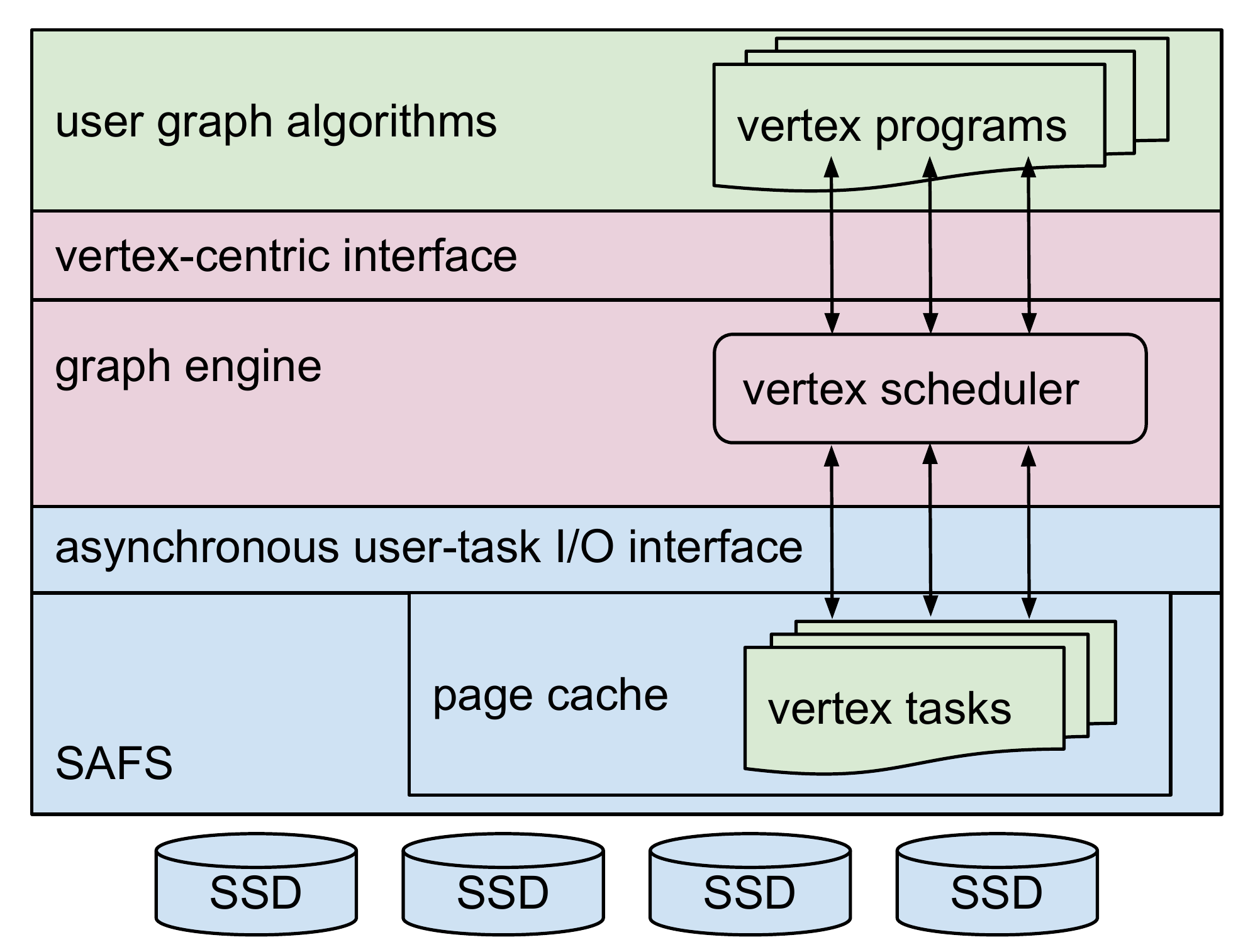}
\vspace{-5pt}
\caption{The architecture of FlashGraph.}
\vspace{-5pt}
\label{arch}
\end{figure}

We build FlashGraph on top of SAFS to fully utilize the high I/O throughput
provided by the SSD array (Figure~\ref{arch}). FlashGraph solely uses
the asynchronous user-task I/O interface of SAFS to reduce the overhead
of accessing data in the page cache, memory consumption, as well as
overlapping computation with I/O. FlashGraph uses
the scalable, lightweight SAFS page cache to buffer the edge lists from SSDs
so that FlashGraph can adapt to applications with different cache hit rates.
%This eliminates the complexity of having another cache in FlashGraph, and
%reduces data redundancy.

A graph algorithm in FlashGraph is composed of many vertex programs that run
inside the graph engine. Each vertex program represents a vertex and has its own
user-defined state and logic. The execution of
vertex programs is subject to scheduling by FlashGraph. When vertex programs
need to access data from SSDs, FlashGraph issues I/O requests to SAFS on behalf
of the vertex programs and pushes part of their computation to SAFS.

\subsection{Execution model}
FlashGraph proceeds in iterations when executing graph algorithms, 
much like other engines. 
In each iteration, FlashGraph processes the vertices activated in the previous
iteration. An algorithm ends when there are no active vertices in the next
iteration.

\begin{figure}[t]
\centering
\includegraphics[scale=0.5]{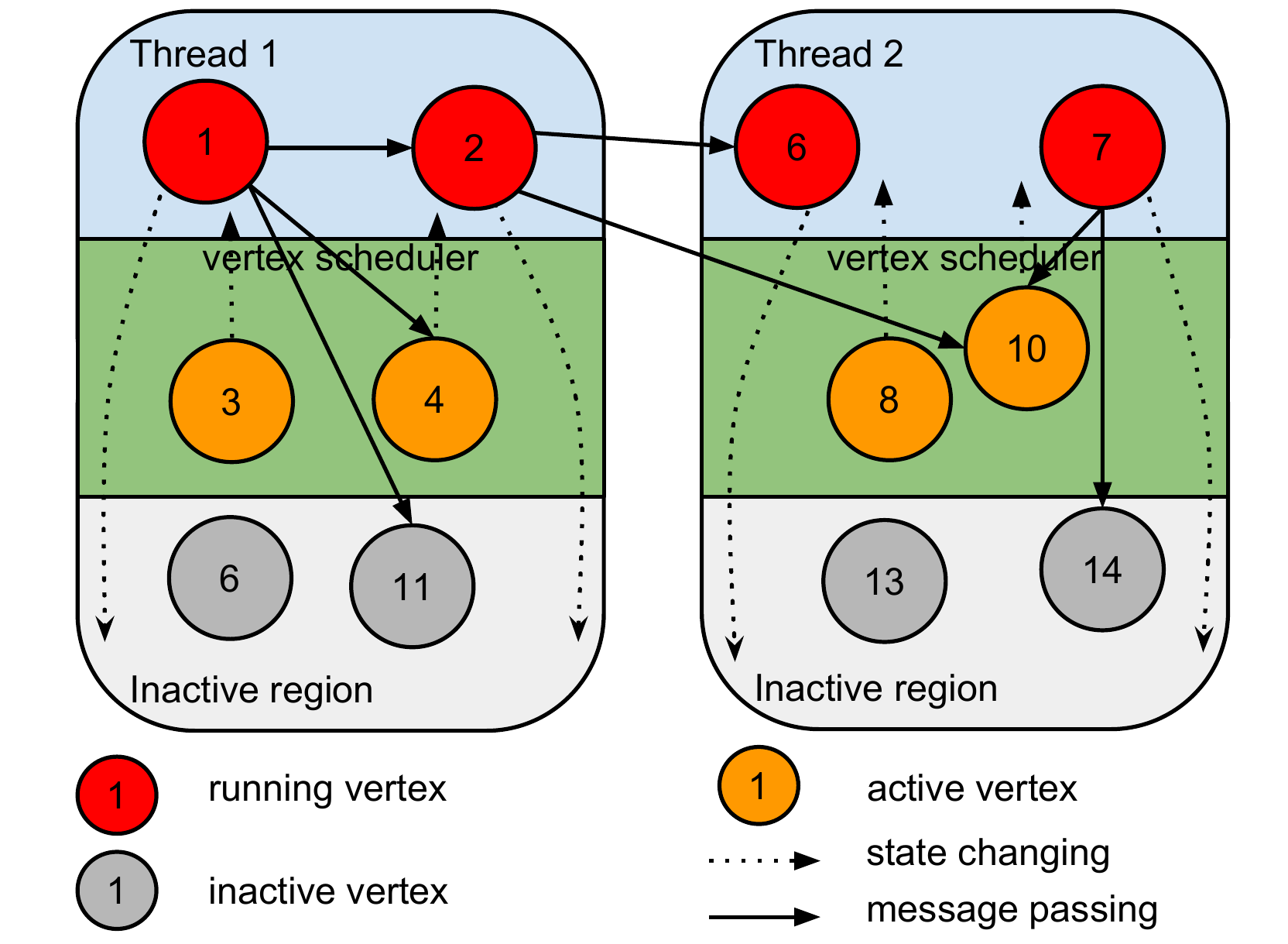}
\vspace{-5pt}
\caption{Execution model in FlashGraph.}
\vspace{-5pt}
\label{exec}
\end{figure}

As shown in Figure \ref{exec},
FlashGraph splits a graph into multiple partitions and assigns a worker thread
to each partition to process vertices. Each worker thread maintains a queue
of active vertices within its own partition and executes user-defined vertex
programs on them. FlashGraph's scheduler both manages the order of execution
of active vertices and guarantees only a fixed number of running vertices
in a thread.

There are three possible states for a vertex: \textit{(i)} running,
\textit{(ii)} active, or \textit{(iii)} inactive.
A vertex can be activated either by other vertices or the graph engine itself.
An active vertex enters the running state when it is scheduled. It remains
in the running state until it finishes its task in the current iteration
and becomes inactive.
The running vertices interact with other vertices via message passing.

\subsection{Programming model} \label{prog_model}
FlashGraph aims at providing a flexible programming interface
to express a variety of graph algorithms and their optimizations.
FlashGraph adopts the vertex-centric programming model commonly used
by other graph engines such as Pregel \cite{pregel} and PowerGraph
\cite{powergraph}. In this programming model, each vertex maintains
vertex state and performs user-defined tasks based on its own state.
A vertex affects the state
of others by sending messages to them as well as activating them. 
% DM commented out: Redundant with sentence in line 295
%Notably,
%FlashGraph allows a vertex to send messages to any vertex in the graph.
We further allow a vertex to read the edge list of any vertex from SSDs.
%as well as the state of any vertex in memory.

The \texttt{run} method (Figure \ref{interface}) is the entry point of
a vertex program in an iteration. It is scheduled and executed exactly
once on each active vertex. It is designed intentionally to have only access
the vertex's own state in this method. A vertex must explicitly request
its own edge list before accessing it because it is common that vertices
are activated but do not perform any computation. Reading a vertex's
edge list by default before executing its \texttt{run} method wastes
I/O bandwidth.

The rest of FlashGraph's programming interface is event-driven to overlap
computation and I/O, and receive notifications from
the graph engine and other vertices. A vertex may receive three types of events:
\begin{itemize}
  \setlength{\itemsep}{0pt}
	\setlength{\parsep}{0pt}
	\setlength{\parskip}{0pt}
	\item  when it receives the edge list of a vertex, FlashGraph executes
		its \texttt{run\_on\_vertex} method.
	\item when it receives a message, FlashGraph executes its
		\texttt{run\_on\_message} method. This method is executed even if
		a vertex is inactive in the iteration.
	\item  when the iteration comes to an end, FlashGraph executes its
		\texttt{run\_on\_iteration\_end} method. A vertex needs to request
		this notification explicitly.
\end{itemize}

\begin{figure}[t]
%\begin{minted}[mathescape,
%		fontsize=\scriptsize,
%		frame=single,
%]{c++}
\begin{lstlisting}
class vertex {
  // entry point (runs in memory)
  void run(graph_engine &g);
  // per vertex computation (runs in the SAFS page cache)
  void run_on_vertex(graph_engine &g, page_vertex &v);
  // process a message (runs in memory)
  void run_on_message(graph_engine &g, vertex_message &msg);
  // run at the end of an iteration when all active vertices
  // in the iteration are processed.
  void run_on_iteration_end(graph_engine &g);
};
\end{lstlisting}
%\end{minted}
\vspace{-5pt}
\caption{The programming interface of FlashGraph.}
\label{interface}
\end{figure}

Given the programming interface, breadth-first search can be simply expressed
as the code in Figure \ref{bfs}. If a vertex has not been visited, it issues
a request to read its edge list in the \texttt{run} method and activates its
neighbors in the \texttt{run\_on\_vertex} method. In this example, vertices do not
receive other events.
\begin{figure}[t]
%\begin{minted}[mathescape,
%		fontsize=\scriptsize,
%		frame=single,
%]{c++}
\begin{lstlisting}
class bfs_vertex: public vertex {
  bool has_visited = false;

  void run(graph_engine &g) {
    if (!has_visited) {
      vertex_id_t id = g.get_vertex_id(*this);
      // Request the edge list of the vertex from SAFS
      request_vertices(&id, 1);
      has_visited = true;
    }
  }

  void run_on_vertex(graph_engine &g, page_vertex &v) {
    vertex_id_t dest_buf[];
    v.read_edges(dest_buf);
    g.activate_vertices(dest_buf, num_dests);
  }
};
\end{lstlisting}
%\end{minted}
\vspace{-5pt}
\caption{Breadth-first search in FlashGraph.}
%\vspace{-10pt}
\label{bfs}
\end{figure}

This interface is designed for better flexibility and gives users fine-grained
programmatic control.
For example, a vertex has to explicitly request its own edge list
so that a graph application can significantly reduce the amount of data brought to memory.
Furthermore, the interface does not constrain the vertices that a vertex
can communicate with or the edge lists that a vertex can request from SSDs.
This flexibility allows FlashGraph to handle algorithms
such as Louvain clustering \cite{louvain}, in which changes to the topology of
the graph occur during computation. It is difficult to express such algorithms
with graph frameworks in which vertices can only interact with direct neighbors.

\subsubsection{Message passing}
%Despite all vertices being on a single machine, we support message passing for
%vertices to push data to other vertices. 
% Better topic sentence than original & avoids the redundancy below
Message passing avoids concurrent data access to the state of other vertices.
A semi-external memory graph engine cannot push data to other vertices by
embedding data on edges like
PowerGraph \cite{powergraph}. Writing data to other vertices directly can cause
race conditions and requires atomic operations or locking for synchronization
on vertex state. Message passing is a light-weight alternative for pushing
data to other vertices. Although message passing requires synchronization to
coordinate messages, it hides explicit synchronization
from users and provides a more user-friendly programming interface.
Furthermore, we can bundle multiple messages in a single packet to reduce
synchronization overhead.

We implement a customized message passing scheme for vertex communication
in FlashGraph. The worker threads send and receive messages on behalf of vertices
and buffer messages to improve performance.
%When a vertex sends a message, the sending thread buffers the message.
%A worker thread maintains a sending buffer for every other thread and it chooses
%a buffer
%for a message based on a graph partition function (Section \ref{partition}).
%A worker thread flushes messages when the buffer gets full. Each worker thread
%maintains a single receiving buffer.
To reduce memory consumption, we process
messages and pass them to vertices when the buffer accumulates a certain
number of messages.

FlashGraph supports multicast to avoid unnecessary message duplication.
It is common that a vertex needs to send the same message to many other vertices.
In this case point-to-point communication causes unnecessary message
duplication. With multicast, FlashGraph simply copies the same message
once to each thread. %However, there is
%some overhead associated with multicast. When a vertex multicasts a message
%to a small number of recipient vertices, FlashGraph uses point-to-point message
%passing. The current implementation only switches to multicast when a vertex
%sends the same message to vertices more than the number of worker threads.
We implement vertex activation with multicast since activation messages contain
no data and are identical. 

%\subsubsection{Synchronous vs. asynchronous computation}
%A side effect of passing messages to vertices during an iteration is that
%FlashGraph supports asynchronous computation. When the state of a vertex
%is changed by a message, the new state can be immediately exposed to other
%vertices in
%the same iteration. It has been demonstrated that asynchronous computation
%has a faster convergence rate than synchronous computation for many graph
%algorithms \cite{para_comp, graphlab}. However, asynchronous computation
%is non-deterministic and some graph algorithms do not converge when they
%run asynchronously.

%FlashGraph provides an additional interface for programmers to enable
%synchronous computation. FlashGraph's approach is similar to Naiad
%\cite{naiad}. It allows a vertex to request a notification
%at the end of an iteration independently. At the end of an iteration,
%FlashGraph invokes \texttt{run\_on\_iteration\_end} on the vertices that
%requested notification during the iteration. To enable synchronous
%computation, each vertex maintains two copies of vertex state: \textit{current state}
%and \textit{future state}. When receiving a message, a vertex only updates
%the future state. At the end of an iteration, the vertex replaces its current
%state with its future state.

\subsection{Data representation in FlashGraph}
FlashGraph uses compact data representations both in memory and on SSDs. A smaller
in-memory data representation allows us to process a larger graph and use a larger
SAFS page cache to improve performance. A smaller data representation on SSDs
allows us to pull more edge lists
from SSDs in the same amount of time, resulting in better performance.

\subsubsection{In-memory data representation}
FlashGraph maintains the following data structures in memory: 
\textit{(i)} a graph index for accessing edge lists on SSDs;
\textit{(ii)} user-defined algorithmic vertex state of all vertices;
\textit{(iii)} vertex status used by FlashGraph;
\textit{(iv)} per-thread message queues.
To save space, we choose to compute some vertex information at runtime, such as
the location of an edge list on SSDs and vertex ID.

The graph index stores a small amount of information for each edge list and compute
their location and size at runtime (Figure \ref{vertex_format}).
Storing both the location and size in memory would require a significant amount
of memory: 12 bytes per vertex in an undirected graph
and 24 bytes in a directed graph. Instead, for almost all vertices, we can
use one byte to store the vertex degree for an undirected vertex and two bytes
for a directed vertex. Knowing the vertex degree, we can compute the edge list size
and further compute their locations, since edge lists on SSDs are sorted by vertex ID.
To balance computation overhead and memory space, we store the locations of a small
number of edge lists in memory. By default, we store one location for every $32$
edge lists, which makes computation overhead almost unnoticeable while the amortized
memory overhead is small. In addition, we store the degree
of large vertices ($\geq 255$) in a hash table. Most real-world graphs follow the power-law
distribution in vertex degree, so there are only a small number of vertices in
the hash table. In our default configuration, each vertex in the index uses slightly
more than $1.25$ bytes in an undirected graph and slightly more than $2.5$ bytes
in a directed graph.

Users define algorithmic vertex state in vertex programs. The semi-external
memory execution model requires the size of vertex state to be a small constant
so FlashGraph can keep it in memory throughout execution. In our experience,
the algorithmic vertex state is usually small. For example, breadth-first search
only needs one byte for each vertex (Figure \ref{bfs}). Many graph algorithms we
implement use no more than eight bytes for each vertex.
Many graph algorithms need to access the vertex ID
that vertex state belongs to in a vertex program. Instead of storing the vertex ID
with vertex state, we compute the vertex ID based on the address of
the vertex state in memory. It is cheap to compute vertex ID most of the time.
It becomes relatively more expensive to compute when FlashGraph starts to balance
load because FlashGraph needs to search multiple partitions for the vertex state
(Section \ref{load_balancing}).

\subsubsection{External-memory data representation}\label{ext_mem}
FlashGraph stores edges and edge attributes of vertices on SSDs.
To amortize the overhead of constructing a graph for analysis in FlashGraph
and reduce SSD wearout, we use a single external-memory data structure for
all graph algorithms supported by FlashGraph.
Since SSDs are still several times slower than RAM, the external-memory data
representation in FlashGraph has to be compact to reduce the amount of data
accessed from SSDs.
%and avoid unnecessary data reads from SSDs.

\begin{figure}[t]
\centering
\includegraphics[scale=0.35]{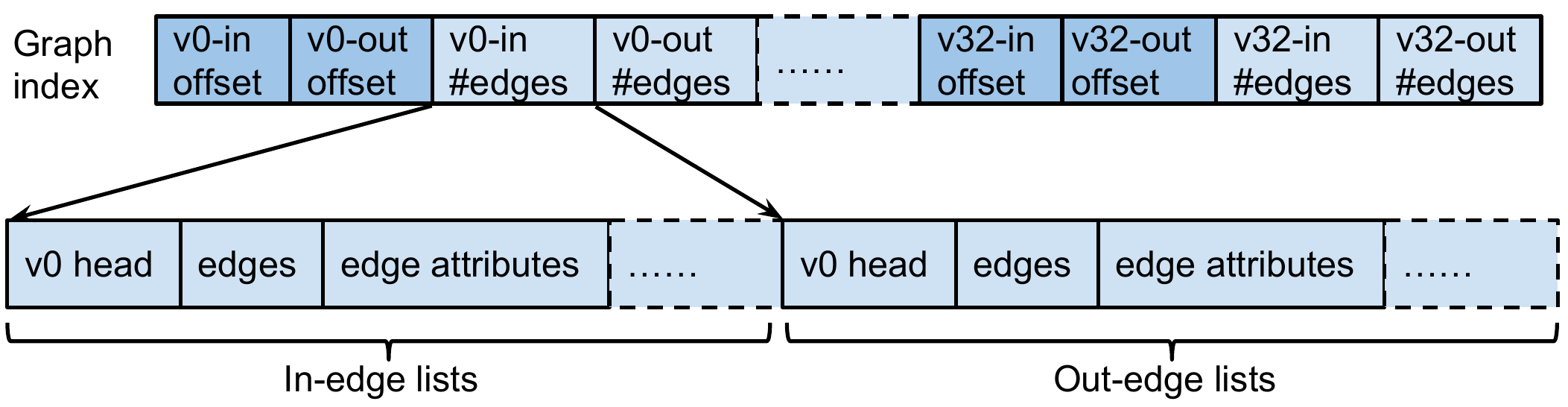}
\caption{The data representation of a directed graph in FlashGraph.
During computation, the graph index is maintained in memory and
the in-edge and out-edge lists are accessed from SSDs.}
\label{vertex_format}
\end{figure}

Figure \ref{vertex_format} shows the data representation of a graph on
SSDs. An edge list has a header, edges and edge attributes.
Edge attributes are stored separately from edges so that
graph applications avoid reading attributes when they are not required.
This strategy is already successfully employed by many database systems
\cite{columnDB}. All of the edge lists stored on SSDs are ordered by
vertex ID, given by the input graph.
%Vertex numbering can greatly affect the performance. A good one increases
%data locality for adjacency list access on SSDs as well as message passing.
%In the future work, we will explore different vertex ordering schemes such
%as shingle ordering \cite{shingle} and SlashBurn \cite{SlashBurn}, or even
%use graph clustering
%schemes such as spectral clustering \cite{spectral} and Louvain clustering
%\cite{louvain} to reorder vertices.

FlashGraph stores the in-edge and out-edge list of a vertex separately
for a directed graph. Many graph applications require only one type of edge.
As such, storing both in-edges and out-edges of a vertex together would cause FlashGraph
to read more data from SSDs. If a graph algorithm does require both in-edges and
out-edges of
vertices, having separate in-edge and out-edge lists could potentially double
the number of I/O requests. However, FlashGraph merges I/O requests
(Section \ref{vertex_access}), which significantly alleviates this problem.

%The full vertex information of a graph and the graph index are stored
%in separate files. FlashGraph uses a single page cache for both files.
%Given the index data is smaller and it is more likely to be used, we tweak
%the page eviction policy and favor to keep index data for a longer time.

%FlashGraph maintains a few bytes of information in memory for each vertex.
%It maintains an array of user-defined vertex states and the vertex IDs.
%In addition,
%it maintains a bitmap indicating the vertices activated in the next iteration
%and a bitmap indicating the stolen vertices during load balancing.
%To represent the active vertices in the current iteration, FlashGraph uses
%either a vertex ID array, if there are only a few active vertices in the
%iteration, or a bitmap, if there are many.

\subsection{Edge list access on SSDs} \label{vertex_access}
% I think we can trim this section. This is a lot of words to say
% 1. FG is general enough to adapt to diverse I/O workloads
% 2. Selectively access edge lists, don't access all like others 3. Merge I/O
%\dz{Needs better topic sentence}
Graph algorithms exhibit varying I/O access patterns in the semi-external
memory computation model. The most prominent is that each vertex accesses
only its own edge list. In this category, graph algorithms such as PageRank
\cite{pagerank} access all edge lists of a graph in an iteration;
graph traversal algorithms require access to many edge lists
in some of their iterations on most real-world graphs. A less common category
of graph algorithms, such as triangle counting, require a vertex to access
the edge lists of many other vertices as well. FlashGraph supports all of these
access patterns and optimizes them differently.

Given the good random I/O performance of SSDs, FlashGraph selectively accesses
the edge lists required by graph algorithms. Most graph algorithms only need to
access a subset of edge lists within an iteration. External-memory graph engines such as
GraphChi \cite{graphchi} and X-Stream \cite{xstream} that sequentially access
\textit{all} edge lists in each
iteration waste I/O bandwidth despite avoiding random I/O access.
Selective access is superior to sequentially accessing the entire graph
in each iteration and significantly reduces the amount of data read from SSDs.
%However, it potentially generates many random I/O accesses to SSDs.

FlashGraph merges I/O requests to maximize its performance.
During an iteration of most algorithms, there are a large number of vertices
that will likely
request many edge lists from SSDs. Given this, it is likely that multiple edge
lists required are stored nearby on SSDs, giving us the opportunity to
merge I/O requests.

FlashGraph globally sorts and merges I/O requests issued by all \textit{active state} vertices
for applications where each vertex requests a single edge list within an iteration.
FlashGraph relies on its vertex scheduler (Section \ref{schedule}) to order all
I/O requests within the iteration. We only merge I/O requests
when they access either the same page or adjacent pages on SSDs.
To minimize the amount of data brought from SSDs, the minimum I/O block
size issued by FlashGraph is one flash page (4KB). As a result,
an I/O request issued by FlashGraph varies from as small as one page to as large as
many megabytes to benefit graph algorithms with various I/O access patterns.

\begin{figure}[t]
\centering
\includegraphics[scale=0.40]{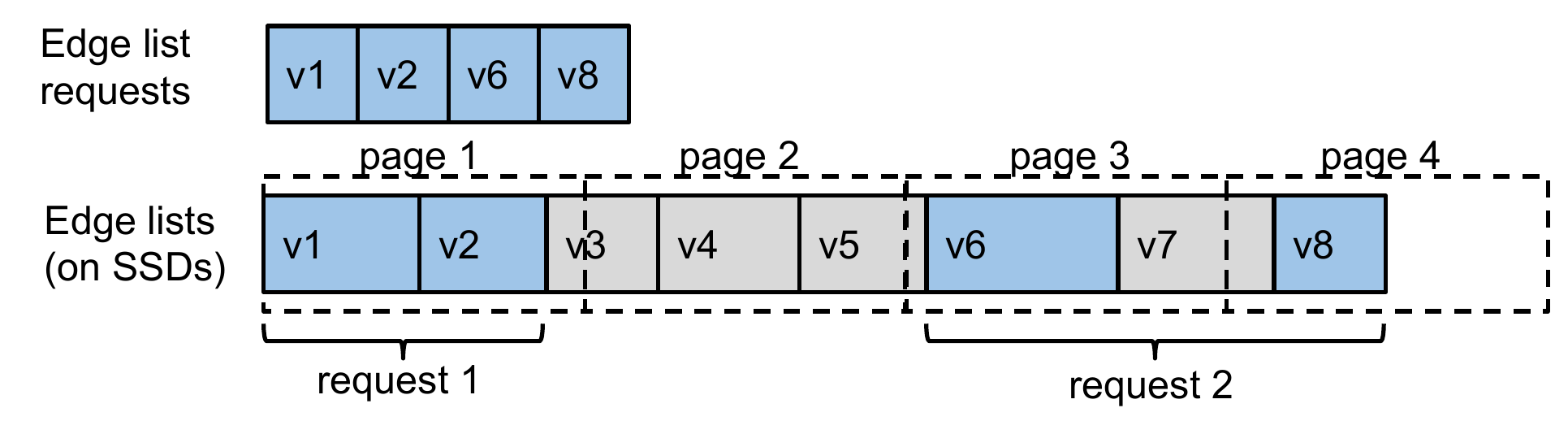}
\caption{FlashGraph accesses edge lists and merges I/O requests.}
\label{fig:vertex_access}
\end{figure}

Figure \ref{fig:vertex_access} illustrates the process of selectively accessing
edge lists on SSDs and merging I/O requests. In this example, the graph algorithm
requests the in-edge lists of four vertices: \texttt{v1, v2, v6 and v8}.
FlashGraph issues I/O requests to access these edge lists from SSDs. Due to our
merging criteria, FlashGraph merges I/O requests for \texttt{v1} and \texttt{v2}
into a single I/O request because they are on the same
page, and merges \texttt{v6} and \texttt{v8} into a single request because
they are on adjacent pages.
As a result, FlashGraph only needs to issue two, as opposed to four, I/O requests to access
four edge lists in this example. %In contrast, a naive implementation would
%result in four I/O accesses to the underlying filesystem.

In the less common case that a vertex requests edge lists of multiple vertices,
FlashGraph must observe I/O requests issued by all \textit{running state} vertices
before sorting them.
In this case, FlashGraph can no longer rely on its
vertex scheduler to reorder I/O requests in an iteration.
The more requests FlashGraph observes, the more likely it is to merge them
and generate cache hits. FlashGraph is only able to observe a relatively
small number of I/O requests, compared to the size of a graph, due to
the memory constraint. It is in this less common case that
FlashGraph relies on SAFS to merge I/O requests to reduce memory consumption.
Finally, to further increase I/O merging and cache hit rates, FlashGraph uses
a flexible vertical graph partitioning scheme (Section \ref{partition}).

%\subsubsection{Vertex prefetching}

\subsection{Vertex scheduling} \label{schedule}
Vertex scheduling greatly affects the performance of graph algorithms.
Intelligent scheduling accelerates the convergence rate and improves I/O
performance. FlashGraph's default scheduler aims to
decrease the number of I/O accesses and increase page cache hit rates.
FlashGraph also allows users to customize the vertex scheduler to optimize
for the I/O access pattern and accelerate the convergence of their algorithms.
For example, scan statistics \cite{scan} in Section \ref{apps}
requires large-degree vertices to be scheduled first to skip expensive computation
on the majority of vertices.

FlashGraph deploys a per-thread vertex scheduler. Each thread schedules vertices
in its own partition independently. This strategy simplifies implementation
and results in framework scalability. The per-thread scheduler keeps multiple
active vertices in the running state so that FlashGraph can observe then
merge many I/O requests issued by vertex programs. In general, FlashGraph
favors a large number of \textit{running state} vertices because it allows
FlashGraph to merge more I/O requests to improve performance. In practice,
performance improvement is no longer noticeable past
$4000$ \textit{running state} vertices per thread.

The default scheduler processes vertices ordered by vertex ID.
This scheduling maximizes merging I/O requests for most graph algorithms
because vertices request their own edge lists in most graph algorithms and
edge lists are ordered by vertex ID on SSDs. For algorithms in which vertex
ordering does not affect the convergence rate, the default scheduler
alternates the direction
that it scans the queue of active vertices between iterations. This
strategy results in pages accessed at the end of the previous iteration
being accessed at the beginning of the current iteration, 
potentially increasing the cache hit rate.

\subsection{Graph partitioning} \label{partition}
\begin{figure}[t]
\centering
\includegraphics[scale=0.5]{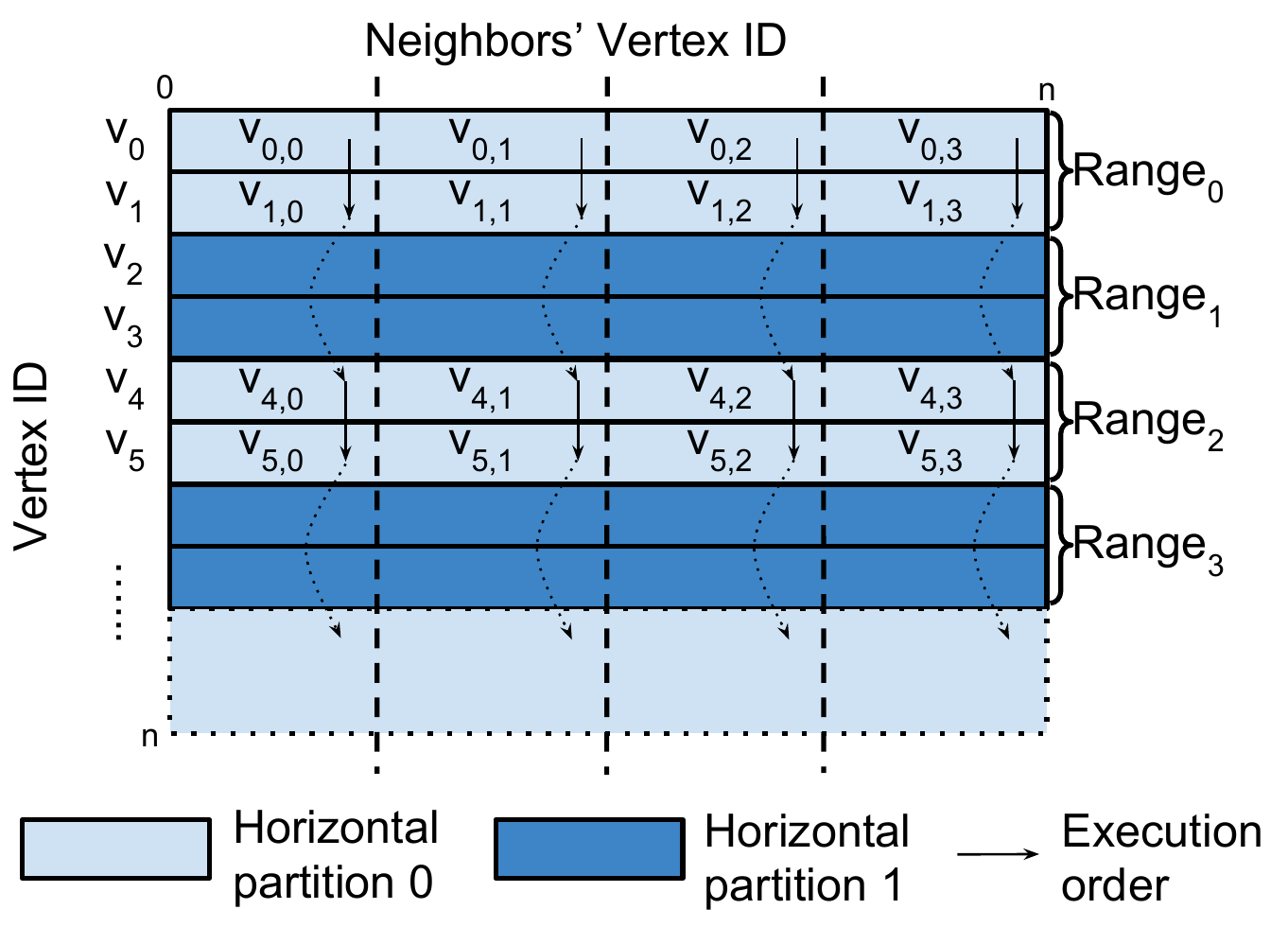}
\caption{An example of 2D partitioning on a graph of $n$ vertices,
	visualized as an adjacency
	matrix. In this example, the graph is split into two horizontal partitions
	and four vertical partitions. The size of a range in a horizontal partition
	is two. $v_{_{i,j}}$ represents vertical partition $j$ of
	vertex $i$. The arrows show the order in which the vertical partitions of
vertices in horizontal partition $0$ are executed in a worker thread.}
\label{graph_part}
\end{figure}

FlashGraph partitions a graph in two dimensions at runtime
(Figure \ref{graph_part}), inspired by two-dimension matrix partitioning.
% Is this meant to be FG assigns each vertex to partition? It reads like
% 1 vertex assigned to multiple patitions.
It assigns each vertex to a partition for parallel processing, shown as \textit{horizontal
partitioning} in Figure \ref{graph_part}. FlashGraph applies the horizontal
partitioning in all graph applications. In addition, it provides a flexible
% is it not an edge-list partitioning scheme at runtime?
runtime edge list partitioning scheme within a horizontal partition, shown
as \textit{vertical partitioning} in Figure \ref{graph_part}. This scheme,
when coupled with the vertex scheduling, can increase
the page cache hit rate for applications that require a vertex to access
the edge lists of many vertices because this increases the possibility that
multiple threads share edge list data
in the cache by accessing the same edge lists concurrently.

FlashGraph assigns a worker thread to each horizontal partition to process
vertices in the partition independently. The worker threads are associated
with specific hardware processors. When a thread processes vertices in its own
partition, all memory accesses to the vertex state are localized to the processor.
As such, our partitioning scheme maximizes data locality in memory access
within each processor.

FlashGraph applies a \textit{range partitioning} function to horizontally
partition a graph.
The function performs a right bit shift on a vertex ID by a predefined number
$r$ and takes the modulo of the shifted result:

%\begin{center}
%\vspace{-8pt}
\indent \indent \indent \texttt{range\_id = vid >> $r$}

\indent \indent \indent \texttt{partition\_id = range\_id \% $n$}
%\vspace{-8pt}
%\end{center}

\noindent As such, a partition consists of multiple vertex ID ranges and
the size of a range is determined by a tunable parameter $r$. $n$ denotes
the number of partitions. All vertices in a partition are assigned to
the same worker thread.

Range partitioning helps FlashGraph to
improve spatial data locality for disk I/O in many graph applications.
FlashGraph uses a per-thread vertex scheduler (Section \ref{schedule}) that
optimizes I/O based on its local knowledge. With range partitioning,
the edge lists of most vertices in the same partition are located adjacently
on SSDs, which helps the per-thread vertex scheduler issue a single large
I/O request to access many edge lists. The range size needs to be at least
as large as the number of vertices being processed in parallel in a thread.
However, a very large range may cause load imbalance because it is difficult
to distribute a small number of ranges to worker threads evenly.
We observe that FlashGraph works well for a graph with over 100 million
vertices when $r$ is between $12$ and $18$.

%Range partitioning also reduces random access to the message buffers for
%message passing. Range partitioning assigns messages sent to
%a certain vertex ID range to the same message buffer. When FlashGraph sends
%a list of messages, sorted by their destinations, it is likely multiple 
%messages that are adjacent in the list will eventually reside in the
%same message buffer. Since the edge lists of vertices on SSDs
%are sorted, it is common that messages sent by a vertex are sorted by their destinations.

%Some algorithms require vertices to read the adjacency lists of multiple
%vertices. In a power-law graph, a vertex may need to access many vertices
%and perform significant computation. Since the load balancer only moves 
%the computation of an entire vertex to other threads, it may be
%insufficient to balance the load of these applications. The access of a large
%number of vertices by a single vertex may also cause cache thrashing,
%which leads to few page cache hits.

The vertical partitioning in FlashGraph allows programmers to split
large vertices into small parts at runtime.
FlashGraph replicates vertex state of vertices that require
vertical partitioning and each copy of the vertex state is referred to
as a \textit{vertex part}. A user has complete freedom to perform
computation on and request edge lists for a \textit{vertex part}.
In an iteration, the default FlashGraph scheduler executes all active
\textit{vertex parts} in the first vertical partition and then proceeds
to the second one and so on.
To avoid concurrent data access to vertex state, a \textit{vertex part}
communicates with other vertices through message passing.

The vertical partitioning increases page cache hits for applications
that require vertices to access the edge lists of their neighbors.
In these applications, a user can partition the edge list of a large
vertex and assign a \textit{vertex part} with part of the edge list.
For example, in Figure~\ref{graph_part}, vertex $v_{_0}$ is split into four
parts: $v_{_{0,0}}$, $v_{_{0,1}}$, $v_{_{0,2}}$ and $v_{_{0,3}}$. Each part $v_{_{0,j}}$
is only responsible for accessing the edge lists of its neighbors with
vertex ID between ${n \over 4} \times j$ and ${n \over 4} \times (j+1)$.
When the scheduler executes \textit{vertex parts} in vertical partition $j$,
only edge lists of vertices with vertex ID between ${n \over 4} \times j$ and
${n \over 4} \times (j+1)$ are accessed from SSDs, thus increasing
the likelihood that an edge list being accessed is in the page cache.

\subsubsection{Load balancing} \label{load_balancing}
FlashGraph provides a dynamic load balancer to address the computational
skew created by high degree vertices in scale-free graphs.
%Since we partition vertices statically, the load balancer ensures that
%we avoid computational imbalance at runtime.
In an iteration, each worker thread processes active vertices in
its own partition. Once a thread finishes processing all active vertices
in its own partition, it `steals' active vertices from other threads and
processes them.
%When it finishes processing a vertex stolen from another worker thread, it
%returns the vertex to its owner thread.
This process continues until no threads have
active vertices left in the current iteration.

%Load balancing only moves computation in the \texttt{run} and
%\texttt{run\_on\_vertex} methods
%to other threads. Message processing always remains in the thread
%where the recipient vertex belongs. This decision gives a simpler
%implementation as well as smaller memory footprint.

Vertical partitioning assists in load balancing. FlashGraph does not execute
computation on a vertex simultaneously in multiple threads to avoid
concurrent data access to the state of a vertex. In the applications where only
a few large vertices dominate the computation of the applications, vertical
partitioning breaks these large vertices into parts so that FlashGraph's load
balancer can move computation of vertex parts to multiple threads,
consequently leading to better load balancing.

\section{Applications} \label{apps}
We evaluate FlashGraph's performance and expressiveness with 
both basic and complex graph algorithms. These algorithms
exhibit different I/O access patterns from the perspective of the framework,
providing a comprehensive evaluation of FlashGraph.
%provides a comprehensive evaluation on the implementation of FlashGraph.

%\vspace{2 mm}
\noindent \textbf{Breadth-first search (BFS)}: It starts with a single active
vertex that activates its neighbors. In each subsequent iteration, the 
active and unvisited vertices activate their neighbors for the next iteration.
The algorithm proceeds
until there are no active vertices left. This requires only out-edge lists.

%BFS is a building block for
%many graph applications such as computing betweenness centrality
%\cite{betweenness} and diameter estimation \cite{diameter}. 
%The algorithm proceeds in iterations. It starts with a single active vertex; 
%this vertex activates its neighbors. In each subsequent iteration, the 
%vertices that are active but have not been visited
%activate their neighbors. The algorithm proceeds until there are no active
%vertices left. Our implementation only uses out-edges to traverse a graph.

%\vspace{2 mm}
%\noindent \textbf {Diameter estimation (DE)}: We compute DE by performing
%a BFS double sweep \cite{diameter}. We use this implementation to estimate
%the diameter of the real-world graphs in Table \ref{graphs}. This requires
%both in-edge and out-edge lists.

%\vspace{2 mm}
\noindent \textbf {Betweenness centrality (BC)}: We compute BC by performing
BFS from a vertex, followed by a back propagation \cite{betweenness}.
For performance evaluation, we perform this process from a single
source vertex. This requires both in-edge and out-edge lists.

%\emph{Diameter estimation (DE)}: Computing the diameter of a graph
%requires one to perform BFS from all vertices in a graph. This is generally
%unrealistic for most large graphs. To compute an estimate, we perform a
%double sweep \cite{diameter}. It performs a BFS from
%a random vertex. When the BFS ends, it performs another BFS from the vertex
%with the longest distance discovered by the previous BFS, and so on.
%This implementation can estimate the diameter of a directed graph with or
%without respecting the edge direction.
%We use this implementation to estimate the diameter of the real-world
%graphs in Table \ref{graphs}.
%Magnien et al. \cite{diameter} describes several
%diameter estimation algorithms. We use a variant of a method called
%double sweep lower bound. 

%\vspace{2 mm}
\noindent \textbf{PageRank (PR)} \cite{pagerank}: In our PR, a vertex
sends the delta of its most recent PR update to its neighbors
who then update thier own PR accordingly \cite{daic}. In PageRank, vertices
converge at different rates. As the algorithm proceeds, fewer and fewer
vertices are activated in an iteration.
We set the maximal number of iterations to 30, matching the
value used by Pregel \cite{pregel}. This requires only out-edge lists.

%\emph{PageRank (PR)} \cite{pagerank}: PageRank is used to
%rank the relevance of a Web page given a query.
%In our implementation, a vertex sends the delta of its most recent %previous
%PageRank update to its neighbors if the delta is larger than
%a threshold. Vertices receive the delta from their neighbors and update
%their own PageRank. In PageRank, vertices converge at different
%rates. As the algorithm proceeds, fewer and fewer vertices are activated
%in an iteration. The algorithm ends when all vertices converge or
%it reaches the maximal number of iterations. We set the maximal number
%of iterations to 30, to match the value used by Pregel \cite{pregel}.

%\vspace{2 mm}
\noindent \textbf{Weakly connected components (WCC)}: WCC in a directed graph
is implemented with label propagation \cite{label_prop}.
All vertices start in their own components, broadcast their component IDs
to all neighbors, and adopt the smallest IDs they observe. A vertex that
does not receive a smaller ID does nothing in the next iteration.
This requires both in-edge and out-edge lists.

%\emph{(Weakly) connected components (WCC)}: Both finding connected components
%in an undirected graph and weakly connected components in a directed graph
%can be implemented with label propagation \cite{label_prop}.
%All vertices start in their own components, broadcast their component IDs
%to all neighbors, and adopt the smallest IDs they observe.
%The algorithm ends when no vertices change their component IDs.

%\vspace{2 mm}
\noindent \textbf{Triangle counting (TC)} \cite{triangle}: A vertex computes
the intersection of its own edge list and the edge list of each neighbor to
look for triangles. We count triangles on only one vertex
in a potential triangle and this vertex then notifies the other two
vertices of the existence of the triangle via message passing. 
This requires both in-edge and out-edge lists.

\noindent \textbf{Scan statistics (SS)} \cite{scan}: The SS metric only requires finding
the maximal locality statistic in the graph, which is the maximal number of edges
in the neighborhood of a vertex. We use a custom FlashGraph user-defined
vertex scheduler that begins computation on vertices with the largest degree first.
With this scheduler, we avoid actual computation for many vertices resulting in a
highly optimized implementation \cite{active_community}. This requires both in-edge
and out-edge lists.

%\vspace{2mm}
These algorithms fall into three categories from the perspective
of I/O access patterns. (1) BFS and betweenness centrality only perform computation
on a subset of vertices in a graph within an iteration, thus they generate
many random I/O accesses. (2) PageRank and (weakly) connected components need
to process all vertices at the beginning, so their I/O access is generally
more sequential.
(3) Triangle counting and scan statistics require a vertex to read many edge
lists. These two graph algorithms are
more I/O intensive than the others and generate many random I/O accesses.

\section{Experimental Evaluation}
We evaluate FlashGraph's performance on the applications in section
\ref{apps} on large real-world graphs. We compare the performance of
FlashGraph with its in-memory implementation as well as other in-memory
graph engines (PowerGraph \cite{powergraph} and Galois \cite{galois}).
For in-memory FlashGraph, we replace SAFS with in-memory data
structures for storing edge lists. %and use an uncompressed index for edge
%lists, which contains the location of each edge list.
We also compare semi-external memory FlashGraph with external-memory
graph engines (GraphChi \cite{graphchi} and X-Stream \cite{xstream}).
We further demonstrate the scalability of FlashGraph on a web graph
of 3.4 billion vertices and 129 billion edges.
We also perform experiments to justify some of our design decisions
that are critical to achieve performance. Throughout all experiments,
we use 32 threads for all graph processing engines.

We conduct all experiments on a non-uniform memory architecture machine with
four Intel Xeon E5-4620 processors, clocked at $2.2$ GHz, and $512$ GB
memory of DDR3-1333. Each processor has eight cores. The machine has three LSI SAS
9207-8e host bus adapters (HBA) connected to a SuperMicro storage chassis,
in which $15$ OCZ Vertex $4$ SSDs are installed. The $15$ SSDs together
deliver approximately $900,000$ reads per second, or around $60,000$ reads
per second per SSD. The machine runs Linux kernel v$3.2.30$.

%\subsection{Graph datasets}
We use the real-world graphs in Table \ref{graphs} for evaluation.
The largest graph is the page graph with $3.4$ billion vertices
and $129$ billion edges. Even the smallest graph we use has $42$ million
vertices and $1.5$ billion edges. The page graph is clustered by domain,
generating good cache hit rates for some graph algorithms.

%We estimate the lower bound of the diameter of these directed graphs via
%our implementation that performs BFS double sweeps \cite{diameter}.
%The estimate ignores edge direction. We observe that the Twitter graph and the
%subdomain Web graph
%have relatively small diameters, while the page graph has a much larger diameter.

\begin{table}
\begin{center}
\footnotesize
\begin{tabular}{|c|c|c|c|c|}
\hline
Graph datasets & \# Vertices & \# Edges & Size & Diameter \\
\hline
Twitter \cite{twitter} & $42$M & $1.5$B & $13$GB & $23$ \\
\hline
Subdomain \cite{web_graph} & $89$M & $2$B & $18$GB & $30$ \\
\hline
Page \cite{web_graph} & $3.4$B & $129$B & $1.1$TB & $650$ \\
%\hline
%RMAT graph \cite{rmat} & & & & \\
\hline
\end{tabular}
\normalsize
\end{center}
\vspace{-10pt}
\caption{Graph data sets. These are directed graphs and the diameter
estimation ignores the edge direction.}
\label{graphs}
\end{table}

\subsection{FlashGraph: in-memory vs. \\
				semi-external memory}
We compare the performance of FlashGraph in semi-external memory with that
of its in-memory implementation to measure the performance loss caused by
accessing edge lists from SSDs.

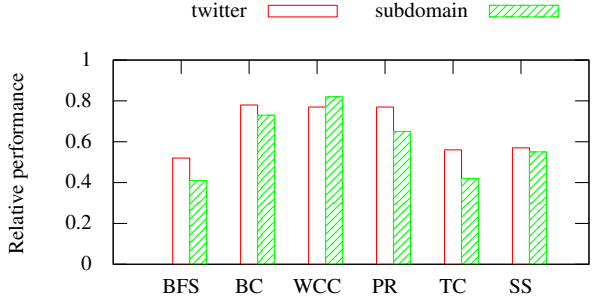
\begin{figure}
	\begin{center}
		\footnotesize
		\vspace{-15pt}
		\begin{tikzpicture}[gnuplot]
%% generated with GNUPLOT 4.6p4 (Lua 5.1; terminal rev. 99, script rev. 100)
%% Wed 31 Dec 2014 12:33:27 AM EST
\path (0.000,0.000) rectangle (8.382,4.318);
\gpcolor{color=gp lt color border}
\gpsetlinetype{gp lt border}
\gpsetlinewidth{1.00}
\draw[gp path] (1.504,0.616)--(1.684,0.616);
\draw[gp path] (7.829,0.616)--(7.649,0.616);
\node[gp node right] at (1.320,0.616) { 0};
\draw[gp path] (1.504,1.159)--(1.684,1.159);
\draw[gp path] (7.829,1.159)--(7.649,1.159);
\node[gp node right] at (1.320,1.159) { 0.2};
\draw[gp path] (1.504,1.703)--(1.684,1.703);
\draw[gp path] (7.829,1.703)--(7.649,1.703);
\node[gp node right] at (1.320,1.703) { 0.4};
\draw[gp path] (1.504,2.246)--(1.684,2.246);
\draw[gp path] (7.829,2.246)--(7.649,2.246);
\node[gp node right] at (1.320,2.246) { 0.6};
\draw[gp path] (1.504,2.790)--(1.684,2.790);
\draw[gp path] (7.829,2.790)--(7.649,2.790);
\node[gp node right] at (1.320,2.790) { 0.8};
\draw[gp path] (1.504,3.333)--(1.684,3.333);
\draw[gp path] (7.829,3.333)--(7.649,3.333);
\node[gp node right] at (1.320,3.333) { 1};
\draw[gp path] (2.408,0.616)--(2.408,0.796);
\draw[gp path] (2.408,3.333)--(2.408,3.153);
\node[gp node center] at (2.408,0.308) {BFS};
\draw[gp path] (3.311,0.616)--(3.311,0.796);
\draw[gp path] (3.311,3.333)--(3.311,3.153);
\node[gp node center] at (3.311,0.308) {BC};
\draw[gp path] (4.215,0.616)--(4.215,0.796);
\draw[gp path] (4.215,3.333)--(4.215,3.153);
\node[gp node center] at (4.215,0.308) {WCC};
\draw[gp path] (5.118,0.616)--(5.118,0.796);
\draw[gp path] (5.118,3.333)--(5.118,3.153);
\node[gp node center] at (5.118,0.308) {PR};
\draw[gp path] (6.022,0.616)--(6.022,0.796);
\draw[gp path] (6.022,3.333)--(6.022,3.153);
\node[gp node center] at (6.022,0.308) {TC};
\draw[gp path] (6.925,0.616)--(6.925,0.796);
\draw[gp path] (6.925,3.333)--(6.925,3.153);
\node[gp node center] at (6.925,0.308) {SS};
\draw[gp path] (1.504,3.333)--(1.504,0.616)--(7.829,0.616)--(7.829,3.333)--cycle;
\node[gp node center,rotate=-270] at (0.246,1.974) {Relative performance};
\node[gp node right] at (3.382,3.984) {twitter};
\def\gpfillpath{(3.566,3.907)--(4.482,3.907)--(4.482,4.061)--(3.566,4.061)--cycle}
\gpfill{color=gpbgfillcolor} \gpfillpath;
\gpfill{color=gp lt color 0,gp pattern 0,pattern color=.} \gpfillpath;
\gpcolor{color=gp lt color 0}
\gpsetlinetype{gp lt plot 0}
\draw[gp path] (3.566,3.907)--(4.482,3.907)--(4.482,4.061)--(3.566,4.061)--cycle;
\def\gpfillpath{(2.295,0.616)--(2.522,0.616)--(2.522,2.030)--(2.295,2.030)--cycle}
\gpfill{color=gpbgfillcolor} \gpfillpath;
\gpfill{color=gp lt color 0,gp pattern 0,pattern color=.} \gpfillpath;
\draw[gp path] (2.295,0.616)--(2.295,2.029)--(2.521,2.029)--(2.521,0.616)--cycle;
\def\gpfillpath{(3.198,0.616)--(3.425,0.616)--(3.425,2.736)--(3.198,2.736)--cycle}
\gpfill{color=gpbgfillcolor} \gpfillpath;
\gpfill{color=gp lt color 0,gp pattern 0,pattern color=.} \gpfillpath;
\draw[gp path] (3.198,0.616)--(3.198,2.735)--(3.424,2.735)--(3.424,0.616)--cycle;
\def\gpfillpath{(4.102,0.616)--(4.329,0.616)--(4.329,2.709)--(4.102,2.709)--cycle}
\gpfill{color=gpbgfillcolor} \gpfillpath;
\gpfill{color=gp lt color 0,gp pattern 0,pattern color=.} \gpfillpath;
\draw[gp path] (4.102,0.616)--(4.102,2.708)--(4.328,2.708)--(4.328,0.616)--cycle;
\def\gpfillpath{(5.005,0.616)--(5.232,0.616)--(5.232,2.709)--(5.005,2.709)--cycle}
\gpfill{color=gpbgfillcolor} \gpfillpath;
\gpfill{color=gp lt color 0,gp pattern 0,pattern color=.} \gpfillpath;
\draw[gp path] (5.005,0.616)--(5.005,2.708)--(5.231,2.708)--(5.231,0.616)--cycle;
\def\gpfillpath{(5.909,0.616)--(6.136,0.616)--(6.136,2.139)--(5.909,2.139)--cycle}
\gpfill{color=gpbgfillcolor} \gpfillpath;
\gpfill{color=gp lt color 0,gp pattern 0,pattern color=.} \gpfillpath;
\draw[gp path] (5.909,0.616)--(5.909,2.138)--(6.135,2.138)--(6.135,0.616)--cycle;
\def\gpfillpath{(6.812,0.616)--(7.039,0.616)--(7.039,2.166)--(6.812,2.166)--cycle}
\gpfill{color=gpbgfillcolor} \gpfillpath;
\gpfill{color=gp lt color 0,gp pattern 0,pattern color=.} \gpfillpath;
\draw[gp path] (6.812,0.616)--(6.812,2.165)--(7.038,2.165)--(7.038,0.616)--cycle;
\gpcolor{color=gp lt color border}
\node[gp node right] at (6.322,3.984) {subdomain};
\def\gpfillpath{(6.506,3.907)--(7.422,3.907)--(7.422,4.061)--(6.506,4.061)--cycle}
\gpfill{color=gpbgfillcolor} \gpfillpath;
\gpfill{color=gp lt color 1,gp pattern 1,pattern color=.} \gpfillpath;
\gpcolor{color=gp lt color 1}
\gpsetlinetype{gp lt plot 1}
\draw[gp path] (6.506,3.907)--(7.422,3.907)--(7.422,4.061)--(6.506,4.061)--cycle;
\def\gpfillpath{(2.521,0.616)--(2.747,0.616)--(2.747,1.731)--(2.521,1.731)--cycle}
\gpfill{color=gpbgfillcolor} \gpfillpath;
\gpfill{color=gp lt color 1,gp pattern 1,pattern color=.} \gpfillpath;
\draw[gp path] (2.521,0.616)--(2.521,1.730)--(2.746,1.730)--(2.746,0.616)--cycle;
\def\gpfillpath{(3.424,0.616)--(3.651,0.616)--(3.651,2.600)--(3.424,2.600)--cycle}
\gpfill{color=gpbgfillcolor} \gpfillpath;
\gpfill{color=gp lt color 1,gp pattern 1,pattern color=.} \gpfillpath;
\draw[gp path] (3.424,0.616)--(3.424,2.599)--(3.650,2.599)--(3.650,0.616)--cycle;
\def\gpfillpath{(4.328,0.616)--(4.555,0.616)--(4.555,2.845)--(4.328,2.845)--cycle}
\gpfill{color=gpbgfillcolor} \gpfillpath;
\gpfill{color=gp lt color 1,gp pattern 1,pattern color=.} \gpfillpath;
\draw[gp path] (4.328,0.616)--(4.328,2.844)--(4.554,2.844)--(4.554,0.616)--cycle;
\def\gpfillpath{(5.231,0.616)--(5.458,0.616)--(5.458,2.383)--(5.231,2.383)--cycle}
\gpfill{color=gpbgfillcolor} \gpfillpath;
\gpfill{color=gp lt color 1,gp pattern 1,pattern color=.} \gpfillpath;
\draw[gp path] (5.231,0.616)--(5.231,2.382)--(5.457,2.382)--(5.457,0.616)--cycle;
\def\gpfillpath{(6.135,0.616)--(6.362,0.616)--(6.362,1.758)--(6.135,1.758)--cycle}
\gpfill{color=gpbgfillcolor} \gpfillpath;
\gpfill{color=gp lt color 1,gp pattern 1,pattern color=.} \gpfillpath;
\draw[gp path] (6.135,0.616)--(6.135,1.757)--(6.361,1.757)--(6.361,0.616)--cycle;
\def\gpfillpath{(7.038,0.616)--(7.265,0.616)--(7.265,2.111)--(7.038,2.111)--cycle}
\gpfill{color=gpbgfillcolor} \gpfillpath;
\gpfill{color=gp lt color 1,gp pattern 1,pattern color=.} \gpfillpath;
\draw[gp path] (7.038,0.616)--(7.038,2.110)--(7.264,2.110)--(7.264,0.616)--cycle;
\gpcolor{color=gp lt color border}
\gpsetlinetype{gp lt border}
\draw[gp path] (1.504,3.333)--(1.504,0.616)--(7.829,0.616)--(7.829,3.333)--cycle;
%% coordinates of the plot area
\gpdefrectangularnode{gp plot 1}{\pgfpoint{1.504cm}{0.616cm}}{\pgfpoint{7.829cm}{3.333cm}}
\end{tikzpicture}
%% gnuplot variables
		\vspace{-15pt}
		\caption{The performance of each application run on semi-external memory
		FlashGraph with 1GB cache relative to in-memory FlashGraph.}
		\label{rt:ext_in}
	\end{center}
\end{figure}

FlashGraph scales by using semi-external memory on SSDs while preserving
up to 80\% performance of its in-memory implementation (Figure \ref{rt:ext_in}).
In this experiment, FlashGraph uses a page cache of 1GB and has low cache hit
rates in most applications. BC, WCC and PR perform the best and have only
small performance degradation when running in external memory.
Even in the worst cases, external-memory BFS and TC realize more than 40\%
performance of their in-memory counterparts on the subdomain Web graph.
%WCC and PageRank generate sequential
%I/O and are bottlenecked by CPU most of the time, so they have small
%performance loss in semi-external memory. BFS generates more
%random I/O access but has little computation, so it is bottlenecked by I/O.
%Betweenness centrality also generates random I/O access but requires more
%computation, so its performance is less bottlenecked by I/O. The performance loss
%for triangle counting comes from random I/O accesses as well as run-time
%vertical partitioning on the graph (Section \ref{partition}); each
%vertex part needs to perform an expensive construction of its algorithmic
%data structure before performing computation.

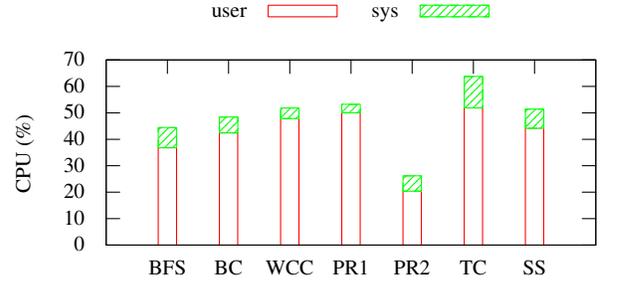
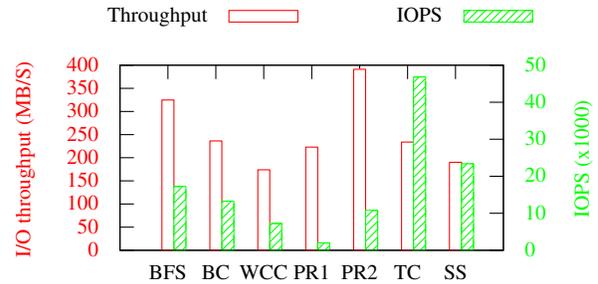
\begin{figure}
\centering
\footnotesize
\vspace{-15pt}
\begin{subfigure}{.5\textwidth}
\begin{tikzpicture}[gnuplot]
%% generated with GNUPLOT 4.6p4 (Lua 5.1; terminal rev. 99, script rev. 100)
%% Fri 02 Jan 2015 11:03:48 AM EST
\path (0.000,0.000) rectangle (8.382,4.064);
\gpcolor{color=gp lt color border}
\gpsetlinetype{gp lt border}
\gpsetlinewidth{1.00}
\draw[gp path] (1.320,0.616)--(1.500,0.616);
\draw[gp path] (7.829,0.616)--(7.649,0.616);
\node[gp node right] at (1.136,0.616) { 0};
\draw[gp path] (1.320,0.968)--(1.500,0.968);
\draw[gp path] (7.829,0.968)--(7.649,0.968);
\node[gp node right] at (1.136,0.968) { 10};
\draw[gp path] (1.320,1.320)--(1.500,1.320);
\draw[gp path] (7.829,1.320)--(7.649,1.320);
\node[gp node right] at (1.136,1.320) { 20};
\draw[gp path] (1.320,1.672)--(1.500,1.672);
\draw[gp path] (7.829,1.672)--(7.649,1.672);
\node[gp node right] at (1.136,1.672) { 30};
\draw[gp path] (1.320,2.023)--(1.500,2.023);
\draw[gp path] (7.829,2.023)--(7.649,2.023);
\node[gp node right] at (1.136,2.023) { 40};
\draw[gp path] (1.320,2.375)--(1.500,2.375);
\draw[gp path] (7.829,2.375)--(7.649,2.375);
\node[gp node right] at (1.136,2.375) { 50};
\draw[gp path] (1.320,2.727)--(1.500,2.727);
\draw[gp path] (7.829,2.727)--(7.649,2.727);
\node[gp node right] at (1.136,2.727) { 60};
\draw[gp path] (1.320,3.079)--(1.500,3.079);
\draw[gp path] (7.829,3.079)--(7.649,3.079);
\node[gp node right] at (1.136,3.079) { 70};
\draw[gp path] (2.134,0.616)--(2.134,0.796);
\draw[gp path] (2.134,3.079)--(2.134,2.899);
\node[gp node center] at (2.134,0.308) {BFS};
\draw[gp path] (2.947,0.616)--(2.947,0.796);
\draw[gp path] (2.947,3.079)--(2.947,2.899);
\node[gp node center] at (2.947,0.308) {BC};
\draw[gp path] (3.761,0.616)--(3.761,0.796);
\draw[gp path] (3.761,3.079)--(3.761,2.899);
\node[gp node center] at (3.761,0.308) {WCC};
\draw[gp path] (4.575,0.616)--(4.575,0.796);
\draw[gp path] (4.575,3.079)--(4.575,2.899);
\node[gp node center] at (4.575,0.308) {PR1};
\draw[gp path] (5.388,0.616)--(5.388,0.796);
\draw[gp path] (5.388,3.079)--(5.388,2.899);
\node[gp node center] at (5.388,0.308) {PR2};
\draw[gp path] (6.202,0.616)--(6.202,0.796);
\draw[gp path] (6.202,3.079)--(6.202,2.899);
\node[gp node center] at (6.202,0.308) {TC};
\draw[gp path] (7.015,0.616)--(7.015,0.796);
\draw[gp path] (7.015,3.079)--(7.015,2.899);
\node[gp node center] at (7.015,0.308) {SS};
\draw[gp path] (1.320,3.079)--(1.320,0.616)--(7.829,0.616)--(7.829,3.079)--cycle;
\node[gp node center,rotate=-270] at (0.246,1.847) {CPU (\%)};
\node[gp node right] at (3.290,3.730) {user};
\def\gpfillpath{(3.474,3.653)--(4.390,3.653)--(4.390,3.807)--(3.474,3.807)--cycle}
\gpfill{color=gpbgfillcolor} \gpfillpath;
\gpfill{color=gp lt color 0,gp pattern 0,pattern color=.} \gpfillpath;
\gpcolor{color=gp lt color 0}
\gpsetlinetype{gp lt plot 0}
\draw[gp path] (3.474,3.653)--(4.390,3.653)--(4.390,3.807)--(3.474,3.807)--cycle;
\def\gpfillpath{(2.012,0.616)--(2.257,0.616)--(2.257,1.912)--(2.012,1.912)--cycle}
\gpfill{color=gpbgfillcolor} \gpfillpath;
\gpfill{color=gp lt color 0,gp pattern 0,pattern color=.} \gpfillpath;
\draw[gp path] (2.012,0.616)--(2.012,1.911)--(2.256,1.911)--(2.256,0.616)--cycle;
\def\gpfillpath{(2.825,0.616)--(3.070,0.616)--(3.070,2.110)--(2.825,2.110)--cycle}
\gpfill{color=gpbgfillcolor} \gpfillpath;
\gpfill{color=gp lt color 0,gp pattern 0,pattern color=.} \gpfillpath;
\draw[gp path] (2.825,0.616)--(2.825,2.109)--(3.069,2.109)--(3.069,0.616)--cycle;
\def\gpfillpath{(3.639,0.616)--(3.884,0.616)--(3.884,2.300)--(3.639,2.300)--cycle}
\gpfill{color=gpbgfillcolor} \gpfillpath;
\gpfill{color=gp lt color 0,gp pattern 0,pattern color=.} \gpfillpath;
\draw[gp path] (3.639,0.616)--(3.639,2.299)--(3.883,2.299)--(3.883,0.616)--cycle;
\def\gpfillpath{(4.452,0.616)--(4.698,0.616)--(4.698,2.377)--(4.452,2.377)--cycle}
\gpfill{color=gpbgfillcolor} \gpfillpath;
\gpfill{color=gp lt color 0,gp pattern 0,pattern color=.} \gpfillpath;
\draw[gp path] (4.452,0.616)--(4.452,2.376)--(4.697,2.376)--(4.697,0.616)--cycle;
\def\gpfillpath{(5.266,0.616)--(5.511,0.616)--(5.511,1.335)--(5.266,1.335)--cycle}
\gpfill{color=gpbgfillcolor} \gpfillpath;
\gpfill{color=gp lt color 0,gp pattern 0,pattern color=.} \gpfillpath;
\draw[gp path] (5.266,0.616)--(5.266,1.334)--(5.510,1.334)--(5.510,0.616)--cycle;
\def\gpfillpath{(6.080,0.616)--(6.325,0.616)--(6.325,2.442)--(6.080,2.442)--cycle}
\gpfill{color=gpbgfillcolor} \gpfillpath;
\gpfill{color=gp lt color 0,gp pattern 0,pattern color=.} \gpfillpath;
\draw[gp path] (6.080,0.616)--(6.080,2.441)--(6.324,2.441)--(6.324,0.616)--cycle;
\def\gpfillpath{(6.893,0.616)--(7.138,0.616)--(7.138,2.172)--(6.893,2.172)--cycle}
\gpfill{color=gpbgfillcolor} \gpfillpath;
\gpfill{color=gp lt color 0,gp pattern 0,pattern color=.} \gpfillpath;
\draw[gp path] (6.893,0.616)--(6.893,2.171)--(7.137,2.171)--(7.137,0.616)--cycle;
\gpcolor{color=gp lt color border}
\node[gp node right] at (5.310,3.730) {sys};
\def\gpfillpath{(5.494,3.653)--(6.410,3.653)--(6.410,3.807)--(5.494,3.807)--cycle}
\gpfill{color=gpbgfillcolor} \gpfillpath;
\gpfill{color=gp lt color 1,gp pattern 1,pattern color=.} \gpfillpath;
\gpcolor{color=gp lt color 1}
\gpsetlinetype{gp lt plot 1}
\draw[gp path] (5.494,3.653)--(6.410,3.653)--(6.410,3.807)--(5.494,3.807)--cycle;
\def\gpfillpath{(2.012,1.911)--(2.257,1.911)--(2.257,2.179)--(2.012,2.179)--cycle}
\gpfill{color=gpbgfillcolor} \gpfillpath;
\gpfill{color=gp lt color 1,gp pattern 1,pattern color=.} \gpfillpath;
\draw[gp path] (2.012,1.911)--(2.012,2.178)--(2.256,2.178)--(2.256,1.911)--cycle;
\def\gpfillpath{(2.825,2.109)--(3.070,2.109)--(3.070,2.320)--(2.825,2.320)--cycle}
\gpfill{color=gpbgfillcolor} \gpfillpath;
\gpfill{color=gp lt color 1,gp pattern 1,pattern color=.} \gpfillpath;
\draw[gp path] (2.825,2.109)--(2.825,2.319)--(3.069,2.319)--(3.069,2.109)--cycle;
\def\gpfillpath{(3.639,2.299)--(3.884,2.299)--(3.884,2.441)--(3.639,2.441)--cycle}
\gpfill{color=gpbgfillcolor} \gpfillpath;
\gpfill{color=gp lt color 1,gp pattern 1,pattern color=.} \gpfillpath;
\draw[gp path] (3.639,2.299)--(3.639,2.440)--(3.883,2.440)--(3.883,2.299)--cycle;
\def\gpfillpath{(4.452,2.376)--(4.698,2.376)--(4.698,2.489)--(4.452,2.489)--cycle}
\gpfill{color=gpbgfillcolor} \gpfillpath;
\gpfill{color=gp lt color 1,gp pattern 1,pattern color=.} \gpfillpath;
\draw[gp path] (4.452,2.376)--(4.452,2.488)--(4.697,2.488)--(4.697,2.376)--cycle;
\def\gpfillpath{(5.266,1.334)--(5.511,1.334)--(5.511,1.537)--(5.266,1.537)--cycle}
\gpfill{color=gpbgfillcolor} \gpfillpath;
\gpfill{color=gp lt color 1,gp pattern 1,pattern color=.} \gpfillpath;
\draw[gp path] (5.266,1.334)--(5.266,1.536)--(5.510,1.536)--(5.510,1.334)--cycle;
\def\gpfillpath{(6.080,2.441)--(6.325,2.441)--(6.325,2.860)--(6.080,2.860)--cycle}
\gpfill{color=gpbgfillcolor} \gpfillpath;
\gpfill{color=gp lt color 1,gp pattern 1,pattern color=.} \gpfillpath;
\draw[gp path] (6.080,2.441)--(6.080,2.859)--(6.324,2.859)--(6.324,2.441)--cycle;
\def\gpfillpath{(6.893,2.171)--(7.138,2.171)--(7.138,2.425)--(6.893,2.425)--cycle}
\gpfill{color=gpbgfillcolor} \gpfillpath;
\gpfill{color=gp lt color 1,gp pattern 1,pattern color=.} \gpfillpath;
\draw[gp path] (6.893,2.171)--(6.893,2.424)--(7.137,2.424)--(7.137,2.171)--cycle;
\gpcolor{color=gp lt color border}
\gpsetlinetype{gp lt border}
\draw[gp path] (1.320,3.079)--(1.320,0.616)--(7.829,0.616)--(7.829,3.079)--cycle;
%% coordinates of the plot area
\gpdefrectangularnode{gp plot 1}{\pgfpoint{1.320cm}{0.616cm}}{\pgfpoint{7.829cm}{3.079cm}}
\end{tikzpicture}
%% gnuplot variables
\label{CPU}
\vspace{-15pt}
\caption{CPU utilization in the user and kernel space. Hyper-threading
enables 64 hardware threads in a 32-core machine, so 50\% CPU utilization
means 32 CPU cores are saturated.}
\end{subfigure}

%\vspace{-5pt}
\begin{subfigure}{.5\textwidth}
\begin{tikzpicture}[gnuplot]
%% generated with GNUPLOT 4.6p4 (Lua 5.1; terminal rev. 99, script rev. 100)
%% Fri 02 Jan 2015 11:03:58 AM EST
\path (0.000,0.000) rectangle (8.382,4.064);
\gpcolor{color=gp lt color border}
\gpsetlinetype{gp lt border}
\gpsetlinewidth{1.00}
\draw[gp path] (1.504,0.616)--(1.684,0.616);
\gpcolor{color=gp lt color 0}
\node[gp node right] at (1.320,0.616) { 0};
\gpcolor{color=gp lt color border}
\draw[gp path] (1.504,0.924)--(1.684,0.924);
\gpcolor{color=gp lt color 0}
\node[gp node right] at (1.320,0.924) { 50};
\gpcolor{color=gp lt color border}
\draw[gp path] (1.504,1.232)--(1.684,1.232);
\gpcolor{color=gp lt color 0}
\node[gp node right] at (1.320,1.232) { 100};
\gpcolor{color=gp lt color border}
\draw[gp path] (1.504,1.540)--(1.684,1.540);
\gpcolor{color=gp lt color 0}
\node[gp node right] at (1.320,1.540) { 150};
\gpcolor{color=gp lt color border}
\draw[gp path] (1.504,1.848)--(1.684,1.848);
\gpcolor{color=gp lt color 0}
\node[gp node right] at (1.320,1.848) { 200};
\gpcolor{color=gp lt color border}
\draw[gp path] (1.504,2.155)--(1.684,2.155);
\gpcolor{color=gp lt color 0}
\node[gp node right] at (1.320,2.155) { 250};
\gpcolor{color=gp lt color border}
\draw[gp path] (1.504,2.463)--(1.684,2.463);
\gpcolor{color=gp lt color 0}
\node[gp node right] at (1.320,2.463) { 300};
\gpcolor{color=gp lt color border}
\draw[gp path] (1.504,2.771)--(1.684,2.771);
\gpcolor{color=gp lt color 0}
\node[gp node right] at (1.320,2.771) { 350};
\gpcolor{color=gp lt color border}
\draw[gp path] (1.504,3.079)--(1.684,3.079);
\gpcolor{color=gp lt color 0}
\node[gp node right] at (1.320,3.079) { 400};
\gpcolor{color=gp lt color border}
\draw[gp path] (2.141,0.616)--(2.141,0.796);
\draw[gp path] (2.141,3.079)--(2.141,2.899);
\node[gp node center] at (2.141,0.308) {BFS};
\draw[gp path] (2.778,0.616)--(2.778,0.796);
\draw[gp path] (2.778,3.079)--(2.778,2.899);
\node[gp node center] at (2.778,0.308) {BC};
\draw[gp path] (3.415,0.616)--(3.415,0.796);
\draw[gp path] (3.415,3.079)--(3.415,2.899);
\node[gp node center] at (3.415,0.308) {WCC};
\draw[gp path] (4.053,0.616)--(4.053,0.796);
\draw[gp path] (4.053,3.079)--(4.053,2.899);
\node[gp node center] at (4.053,0.308) {PR1};
\draw[gp path] (4.690,0.616)--(4.690,0.796);
\draw[gp path] (4.690,3.079)--(4.690,2.899);
\node[gp node center] at (4.690,0.308) {PR2};
\draw[gp path] (5.327,0.616)--(5.327,0.796);
\draw[gp path] (5.327,3.079)--(5.327,2.899);
\node[gp node center] at (5.327,0.308) {TC};
\draw[gp path] (5.964,0.616)--(5.964,0.796);
\draw[gp path] (5.964,3.079)--(5.964,2.899);
\node[gp node center] at (5.964,0.308) {SS};
\draw[gp path] (6.601,0.616)--(6.421,0.616);
\gpcolor{color=gp lt color 1}
\node[gp node left] at (6.785,0.616) { 0};
\gpcolor{color=gp lt color border}
\draw[gp path] (6.601,1.109)--(6.421,1.109);
\gpcolor{color=gp lt color 1}
\node[gp node left] at (6.785,1.109) { 10};
\gpcolor{color=gp lt color border}
\draw[gp path] (6.601,1.601)--(6.421,1.601);
\gpcolor{color=gp lt color 1}
\node[gp node left] at (6.785,1.601) { 20};
\gpcolor{color=gp lt color border}
\draw[gp path] (6.601,2.094)--(6.421,2.094);
\gpcolor{color=gp lt color 1}
\node[gp node left] at (6.785,2.094) { 30};
\gpcolor{color=gp lt color border}
\draw[gp path] (6.601,2.586)--(6.421,2.586);
\gpcolor{color=gp lt color 1}
\node[gp node left] at (6.785,2.586) { 40};
\gpcolor{color=gp lt color border}
\draw[gp path] (6.601,3.079)--(6.421,3.079);
\gpcolor{color=gp lt color 1}
\node[gp node left] at (6.785,3.079) { 50};
\gpcolor{color=gp lt color border}
\draw[gp path] (1.504,3.079)--(1.504,0.616)--(6.601,0.616)--(6.601,3.079)--cycle;
\gpcolor{color=gp lt color 0}
\node[gp node center,rotate=-270] at (0.246,1.847) {I/O throughput (MB/S)};
\gpcolor{color=gp lt color 1}
\node[gp node center,rotate=-270] at (7.674,1.847) {IOPS (x1000)};
\gpcolor{color=gp lt color border}
\node[gp node right] at (2.768,3.730) {Throughput};
\def\gpfillpath{(2.952,3.653)--(3.868,3.653)--(3.868,3.807)--(2.952,3.807)--cycle}
\gpfill{color=gpbgfillcolor} \gpfillpath;
\gpfill{color=gp lt color 0,gp pattern 0,pattern color=.} \gpfillpath;
\gpcolor{color=gp lt color 0}
\gpsetlinetype{gp lt plot 0}
\draw[gp path] (2.952,3.653)--(3.868,3.653)--(3.868,3.807)--(2.952,3.807)--cycle;
\def\gpfillpath{(2.061,0.616)--(2.222,0.616)--(2.222,2.617)--(2.061,2.617)--cycle}
\gpfill{color=gpbgfillcolor} \gpfillpath;
\gpfill{color=gp lt color 0,gp pattern 0,pattern color=.} \gpfillpath;
\draw[gp path] (2.061,0.616)--(2.061,2.616)--(2.221,2.616)--(2.221,0.616)--cycle;
\def\gpfillpath{(2.699,0.616)--(2.859,0.616)--(2.859,2.072)--(2.699,2.072)--cycle}
\gpfill{color=gpbgfillcolor} \gpfillpath;
\gpfill{color=gp lt color 0,gp pattern 0,pattern color=.} \gpfillpath;
\draw[gp path] (2.699,0.616)--(2.699,2.071)--(2.858,2.071)--(2.858,0.616)--cycle;
\def\gpfillpath{(3.336,0.616)--(3.496,0.616)--(3.496,1.688)--(3.336,1.688)--cycle}
\gpfill{color=gpbgfillcolor} \gpfillpath;
\gpfill{color=gp lt color 0,gp pattern 0,pattern color=.} \gpfillpath;
\draw[gp path] (3.336,0.616)--(3.336,1.687)--(3.495,1.687)--(3.495,0.616)--cycle;
\def\gpfillpath{(3.973,0.616)--(4.133,0.616)--(4.133,1.990)--(3.973,1.990)--cycle}
\gpfill{color=gpbgfillcolor} \gpfillpath;
\gpfill{color=gp lt color 0,gp pattern 0,pattern color=.} \gpfillpath;
\draw[gp path] (3.973,0.616)--(3.973,1.989)--(4.132,1.989)--(4.132,0.616)--cycle;
\def\gpfillpath{(4.610,0.616)--(4.770,0.616)--(4.770,3.025)--(4.610,3.025)--cycle}
\gpfill{color=gpbgfillcolor} \gpfillpath;
\gpfill{color=gp lt color 0,gp pattern 0,pattern color=.} \gpfillpath;
\draw[gp path] (4.610,0.616)--(4.610,3.024)--(4.769,3.024)--(4.769,0.616)--cycle;
\def\gpfillpath{(5.247,0.616)--(5.407,0.616)--(5.407,2.057)--(5.247,2.057)--cycle}
\gpfill{color=gpbgfillcolor} \gpfillpath;
\gpfill{color=gp lt color 0,gp pattern 0,pattern color=.} \gpfillpath;
\draw[gp path] (5.247,0.616)--(5.247,2.056)--(5.406,2.056)--(5.406,0.616)--cycle;
\def\gpfillpath{(5.884,0.616)--(6.045,0.616)--(6.045,1.788)--(5.884,1.788)--cycle}
\gpfill{color=gpbgfillcolor} \gpfillpath;
\gpfill{color=gp lt color 0,gp pattern 0,pattern color=.} \gpfillpath;
\draw[gp path] (5.884,0.616)--(5.884,1.787)--(6.044,1.787)--(6.044,0.616)--cycle;
\gpcolor{color=gp lt color border}
\node[gp node right] at (5.892,3.730) {IOPS};
\def\gpfillpath{(6.076,3.653)--(6.992,3.653)--(6.992,3.807)--(6.076,3.807)--cycle}
\gpfill{color=gpbgfillcolor} \gpfillpath;
\gpfill{color=gp lt color 1,gp pattern 1,pattern color=.} \gpfillpath;
\gpcolor{color=gp lt color 1}
\gpsetlinetype{gp lt plot 1}
\draw[gp path] (6.076,3.653)--(6.992,3.653)--(6.992,3.807)--(6.076,3.807)--cycle;
\def\gpfillpath{(2.221,0.616)--(2.381,0.616)--(2.381,1.466)--(2.221,1.466)--cycle}
\gpfill{color=gpbgfillcolor} \gpfillpath;
\gpfill{color=gp lt color 1,gp pattern 1,pattern color=.} \gpfillpath;
\draw[gp path] (2.221,0.616)--(2.221,1.465)--(2.380,1.465)--(2.380,0.616)--cycle;
\def\gpfillpath{(2.858,0.616)--(3.018,0.616)--(3.018,1.270)--(2.858,1.270)--cycle}
\gpfill{color=gpbgfillcolor} \gpfillpath;
\gpfill{color=gp lt color 1,gp pattern 1,pattern color=.} \gpfillpath;
\draw[gp path] (2.858,0.616)--(2.858,1.269)--(3.017,1.269)--(3.017,0.616)--cycle;
\def\gpfillpath{(3.495,0.616)--(3.655,0.616)--(3.655,0.973)--(3.495,0.973)--cycle}
\gpfill{color=gpbgfillcolor} \gpfillpath;
\gpfill{color=gp lt color 1,gp pattern 1,pattern color=.} \gpfillpath;
\draw[gp path] (3.495,0.616)--(3.495,0.972)--(3.654,0.972)--(3.654,0.616)--cycle;
\def\gpfillpath{(4.132,0.616)--(4.292,0.616)--(4.292,0.714)--(4.132,0.714)--cycle}
\gpfill{color=gpbgfillcolor} \gpfillpath;
\gpfill{color=gp lt color 1,gp pattern 1,pattern color=.} \gpfillpath;
\draw[gp path] (4.132,0.616)--(4.132,0.713)--(4.291,0.713)--(4.291,0.616)--cycle;
\def\gpfillpath{(4.769,0.616)--(4.930,0.616)--(4.930,1.148)--(4.769,1.148)--cycle}
\gpfill{color=gpbgfillcolor} \gpfillpath;
\gpfill{color=gp lt color 1,gp pattern 1,pattern color=.} \gpfillpath;
\draw[gp path] (4.769,0.616)--(4.769,1.147)--(4.929,1.147)--(4.929,0.616)--cycle;
\def\gpfillpath{(5.406,0.616)--(5.567,0.616)--(5.567,2.923)--(5.406,2.923)--cycle}
\gpfill{color=gpbgfillcolor} \gpfillpath;
\gpfill{color=gp lt color 1,gp pattern 1,pattern color=.} \gpfillpath;
\draw[gp path] (5.406,0.616)--(5.406,2.922)--(5.566,2.922)--(5.566,0.616)--cycle;
\def\gpfillpath{(6.044,0.616)--(6.204,0.616)--(6.204,1.768)--(6.044,1.768)--cycle}
\gpfill{color=gpbgfillcolor} \gpfillpath;
\gpfill{color=gp lt color 1,gp pattern 1,pattern color=.} \gpfillpath;
\draw[gp path] (6.044,0.616)--(6.044,1.767)--(6.203,1.767)--(6.203,0.616)--cycle;
\gpcolor{color=gp lt color border}
\gpsetlinetype{gp lt border}
\draw[gp path] (1.504,3.079)--(1.504,0.616)--(6.601,0.616)--(6.601,3.079)--cycle;
%% coordinates of the plot area
\gpdefrectangularnode{gp plot 1}{\pgfpoint{1.504cm}{0.616cm}}{\pgfpoint{6.601cm}{3.079cm}}
\end{tikzpicture}
%% gnuplot variables
\label{IO}
\vspace{-15pt}
\caption{I/O utilization.}
\end{subfigure}
\caption{CPU and I/O utilization of FlashGraph on the subdomain Web graph.
PR1 is the first 15 iterations of PageRank and PR2 is the last 15 iterations
of PageRank.}
\label{CPU_IO}
\end{figure}

Given around a million IOPS from the SSD array, we observe that most applications
saturate CPU before saturating I/O.
Figure \ref{CPU_IO} shows the CPU and I/O utilization of our applications
in semi-external memory on the subdomain Web graph. Our machine has hyper-threading
enabled, which results in 64 hardware threads in a 32-core machine,
so 32 CPU cores are actually saturated when the CPU utilization gets to 50\%.
Both PageRank and WCC have very sequential I/O and are completely bottlenecked
by the CPU at the beginning. Triangle counting saturates both CPU and I/O.
It generates many small I/O requests and
consumes considerable CPU time in the kernel space (almost 8 CPU cores).
BFS generates very high I/O throughput in terms of bytes per second
but has low CPU utilization, which suggests BFS is most likely bottlenecked by
I/O. Although betweenness centrality has exactly the same I/O access pattern
as BFS, it has lower I/O throughput and higher CPU utilization because it
requires more computation than BFS. As a result, betweenness centrality is
bottlenecked by CPU most of the time. The CPU-bound applications tend to have
a small performance gap between in-memory and semi-external memory implementations.

%\dz{We should include load time.}

\subsection{FlashGraph vs. in-memory engines}
We compare the performance of FlashGraph to PowerGraph \cite{powergraph},
a popular distributed
in-memory graph engine, and Galois \cite{galois}, a state-of-art
in-memory graph engine. FlashGraph and Powergraph provide a general high-level
vertex-centric programming interface, whereas Galois provides a low-level
programming abstraction for building graph engines.
%, so programming graph applications on Galois
%directly avoids overhead from vertex-centric programming interface.
We run these three graph engines on the Twitter and subdomain Web graphs.
Unfortunately, the Web page graph is too large for in-memory graph engines.
We run PowerGraph in multithread mode to
achieve its best performance and use its synchronous execution engine
because it performs better than the asynchronous one on both graphs.

%For a complete comparison, we implement the same algorithm for computing
%scan statistics in PowerGraph as the one in FlashGraph. The implementation
%requires the use of a PowerGraph asynchronous engine to perform
%pruning. However, PowerGraph does not allow users to customize scheduling,
%so the pruning optimization is less effective than in FlashGraph.

%\begin{table}
%\begin{center}
%\footnotesize
%\begin{tabular}{|l|l|r|r|r|r|}
%\hline
%Graph & Alg & FG-mem & FG-$1$ GB & PG & Galois \\
%\hline
%Twitter & BFS & $1.85$ & $3.56$ & $10.20$ & $\mathbf{1.26}$ \\ \cline{2-6}
%& BC & & & & $7.6$ \\ \cline{2-6}
%& WCC & $\mathbf{6.08}$ & $7.91$ & $106.10$ & $7.72$ \\ \cline{2-6}
%& PR & $\mathbf{41.60}$ & $54.07$ & $447.60$ & $80.42$ \\ \cline{2-6}
%& TC & $\mathbf{299.39}$ & $532.21$ & $1309.87$ & N/A \\ \cline{2-6}
%& SS & $\mathbf{128.79}$ & $224.99$ & $1090.17$ & N/A \\ \cline{2-6}
%\hline
%Subdomain & BFS & $2.80$ & $6.87$ & $68.50$ & $\mathbf{1.82}$ \\ \cline{2-6}
%& BC & & & & $10.60$ \\ \cline{2-6}
%& WCC & $\mathbf{12.07}$ & $14.69$ & $238.80$ & $12.20$ \\ \cline{2-6}
%& PR & $\mathbf{38.38}$ & $59.46$ & $741.90$ & $137.00$ \\ \cline{2-6}
%& TC & $\mathbf{127.81}$ & $320.74$ & $300.97$ & N/A \\ \cline{2-6}
%& SS & $\mathbf{20.26}$ & $36.52$ & $1692.53$ & N/A \\ \cline{2-6}
%\hline
%\end{tabular}
%\normalsize
%\end{center}
%\caption{The runtime (sec) of the applications in FlashGraph of both
%	in-memory version and semi-external memory version, PowerGraph
%	and Galois on the Twitter graph and the subdomain graph.}
%\label{vs_powergraph}
%\end{table}

\begin{figure}[thb!]
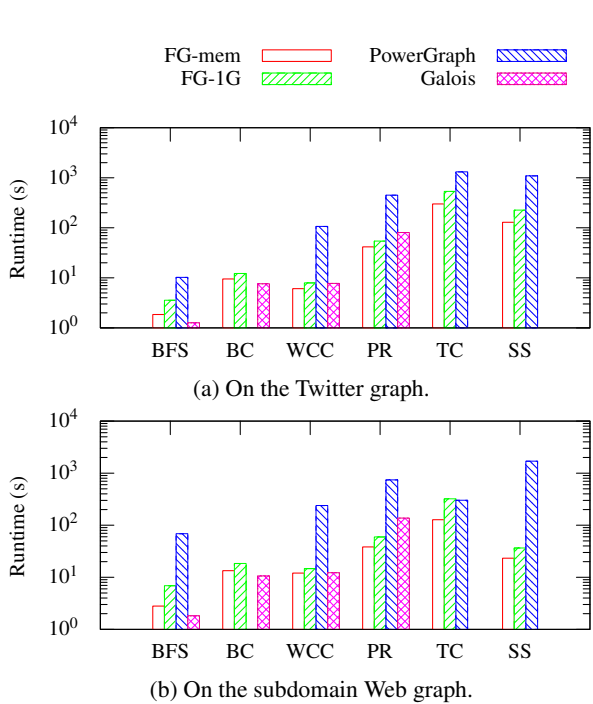

\centering
\footnotesize
%\vspace{-15pt}
\begin{subfigure}{.5\textwidth}
\include{comp.in.mem.twitter}
\label{rt:twitter}
\vspace{-15pt}
\caption{On the Twitter graph.}
\end{subfigure}

\vspace{-15pt}
\begin{subfigure}{.5\textwidth}
\include{comp.in.mem.subdomain}
\label{rt:subdomain}
\vspace{-15pt}
\caption{On the subdomain Web graph.}
\end{subfigure}
\caption{The runtime of different graph engines. FG-mem is in-memory FlashGraph.
FG-1G is semi-external memory FlashGraph with a page cache of $1$ GB.}
\label{rt:in_mem}
\end{figure}

Both in-memory and semi-external memory FlashGraph performs comparably
to Galois, while significantly outperforming PowerGraph (Figure
\ref{rt:in_mem}).
In-memory FlashGraph outperforms Galois in WCC and PageRank.
It performs worse than Galois in graph traversal applications such
as BFS and betweenness centrality, because Galois uses
a different algorithm \cite{fast_bfs} for BFS. The algorithm
reduces the number of edges traversed in both applications. The same
algorithm could be implemented in FlashGraph but would not benefit
semi-external memory FlashGraph because the algorithm requires access to
both in-edge and out-edge lists, thus, significantly increasing the amount
of data read from SSDs.

\subsection{FlashGraph vs. external memory \\ engines}
We compare the performance of FlashGraph to that of two external-memory graph engines,
X-Stream \cite{xstream} and GraphChi \cite{graphchi}.
We run FlashGraph in semi-external memory and use a $1$ GB page cache.
We construct a software RAID on the same SSD array to run X-Stream and GraphChi.
%Existing implementations of PageRank within distributions of both X-Stream
%and GraphChi cannot converge by themselves, hence we 
%set a hard limit of $30$ iterations for all frameworks.
Note that GraphChi does not provide a BFS implementation, and
X-Stream implements triangle counting via a semi-streaming algorithm
\cite{Semi_streaming}.

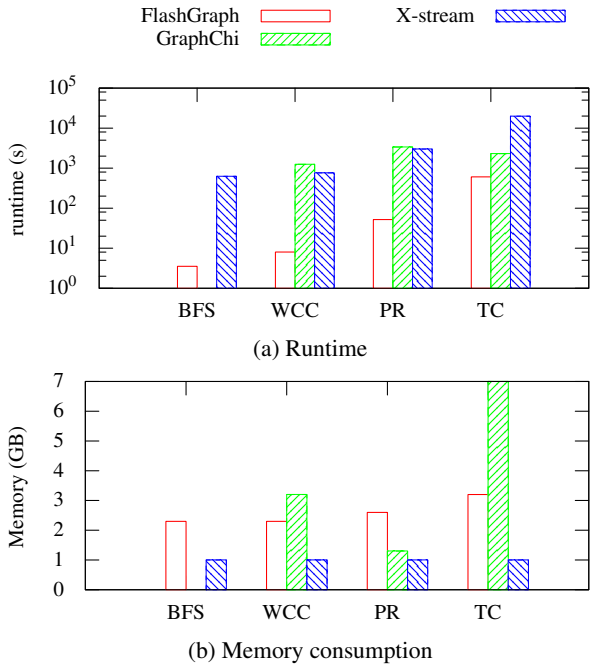
\begin{figure}[t]
\centering
\footnotesize
\vspace{-15pt}
\begin{subfigure}{.5\textwidth}
\begin{tikzpicture}[gnuplot]
%% generated with GNUPLOT 4.6p4 (Lua 5.1; terminal rev. 99, script rev. 100)
%% Tue 23 Sep 2014 01:09:03 PM EDT
\path (0.000,0.000) rectangle (8.382,4.572);
\gpcolor{color=gp lt color border}
\gpsetlinetype{gp lt border}
\gpsetlinewidth{1.00}
\draw[gp path] (1.320,0.616)--(1.500,0.616);
\draw[gp path] (7.829,0.616)--(7.649,0.616);
\node[gp node right] at (1.136,0.616) {$10^{0}$};
\draw[gp path] (1.320,0.776)--(1.410,0.776);
\draw[gp path] (7.829,0.776)--(7.739,0.776);
\draw[gp path] (1.320,0.988)--(1.410,0.988);
\draw[gp path] (7.829,0.988)--(7.739,0.988);
\draw[gp path] (1.320,1.097)--(1.410,1.097);
\draw[gp path] (7.829,1.097)--(7.739,1.097);
\draw[gp path] (1.320,1.149)--(1.500,1.149);
\draw[gp path] (7.829,1.149)--(7.649,1.149);
\node[gp node right] at (1.136,1.149) {$10^{1}$};
\draw[gp path] (1.320,1.309)--(1.410,1.309);
\draw[gp path] (7.829,1.309)--(7.739,1.309);
\draw[gp path] (1.320,1.521)--(1.410,1.521);
\draw[gp path] (7.829,1.521)--(7.739,1.521);
\draw[gp path] (1.320,1.630)--(1.410,1.630);
\draw[gp path] (7.829,1.630)--(7.739,1.630);
\draw[gp path] (1.320,1.681)--(1.500,1.681);
\draw[gp path] (7.829,1.681)--(7.649,1.681);
\node[gp node right] at (1.136,1.681) {$10^{2}$};
\draw[gp path] (1.320,1.842)--(1.410,1.842);
\draw[gp path] (7.829,1.842)--(7.739,1.842);
\draw[gp path] (1.320,2.053)--(1.410,2.053);
\draw[gp path] (7.829,2.053)--(7.739,2.053);
\draw[gp path] (1.320,2.162)--(1.410,2.162);
\draw[gp path] (7.829,2.162)--(7.739,2.162);
\draw[gp path] (1.320,2.214)--(1.500,2.214);
\draw[gp path] (7.829,2.214)--(7.649,2.214);
\node[gp node right] at (1.136,2.214) {$10^{3}$};
\draw[gp path] (1.320,2.374)--(1.410,2.374);
\draw[gp path] (7.829,2.374)--(7.739,2.374);
\draw[gp path] (1.320,2.586)--(1.410,2.586);
\draw[gp path] (7.829,2.586)--(7.739,2.586);
\draw[gp path] (1.320,2.695)--(1.410,2.695);
\draw[gp path] (7.829,2.695)--(7.739,2.695);
\draw[gp path] (1.320,2.746)--(1.500,2.746);
\draw[gp path] (7.829,2.746)--(7.649,2.746);
\node[gp node right] at (1.136,2.746) {$10^{4}$};
\draw[gp path] (1.320,2.907)--(1.410,2.907);
\draw[gp path] (7.829,2.907)--(7.739,2.907);
\draw[gp path] (1.320,3.119)--(1.410,3.119);
\draw[gp path] (7.829,3.119)--(7.739,3.119);
\draw[gp path] (1.320,3.227)--(1.410,3.227);
\draw[gp path] (7.829,3.227)--(7.739,3.227);
\draw[gp path] (1.320,3.279)--(1.500,3.279);
\draw[gp path] (7.829,3.279)--(7.649,3.279);
\node[gp node right] at (1.136,3.279) {$10^{5}$};
\draw[gp path] (2.622,0.616)--(2.622,0.796);
\draw[gp path] (2.622,3.279)--(2.622,3.099);
\node[gp node center] at (2.622,0.308) {BFS};
\draw[gp path] (3.924,0.616)--(3.924,0.796);
\draw[gp path] (3.924,3.279)--(3.924,3.099);
\node[gp node center] at (3.924,0.308) {WCC};
\draw[gp path] (5.225,0.616)--(5.225,0.796);
\draw[gp path] (5.225,3.279)--(5.225,3.099);
\node[gp node center] at (5.225,0.308) {PR};
\draw[gp path] (6.527,0.616)--(6.527,0.796);
\draw[gp path] (6.527,3.279)--(6.527,3.099);
\node[gp node center] at (6.527,0.308) {TC};
\draw[gp path] (1.320,3.279)--(1.320,0.616)--(7.829,0.616)--(7.829,3.279)--cycle;
\node[gp node center,rotate=-270] at (0.246,1.947) {runtime (s)};
\node[gp node right] at (3.290,4.238) {FlashGraph};
\def\gpfillpath{(3.474,4.161)--(4.390,4.161)--(4.390,4.315)--(3.474,4.315)--cycle}
\gpfill{color=gpbgfillcolor} \gpfillpath;
\gpfill{color=gp lt color 0,gp pattern 0,pattern color=.} \gpfillpath;
\gpcolor{color=gp lt color 0}
\gpsetlinetype{gp lt plot 0}
\draw[gp path] (3.474,4.161)--(4.390,4.161)--(4.390,4.315)--(3.474,4.315)--cycle;
\def\gpfillpath{(2.361,0.616)--(2.623,0.616)--(2.623,0.909)--(2.361,0.909)--cycle}
\gpfill{color=gpbgfillcolor} \gpfillpath;
\gpfill{color=gp lt color 0,gp pattern 0,pattern color=.} \gpfillpath;
\draw[gp path] (2.361,0.616)--(2.361,0.908)--(2.622,0.908)--(2.622,0.616)--cycle;
\def\gpfillpath{(3.663,0.616)--(3.925,0.616)--(3.925,1.099)--(3.663,1.099)--cycle}
\gpfill{color=gpbgfillcolor} \gpfillpath;
\gpfill{color=gp lt color 0,gp pattern 0,pattern color=.} \gpfillpath;
\draw[gp path] (3.663,0.616)--(3.663,1.098)--(3.924,1.098)--(3.924,0.616)--cycle;
\def\gpfillpath{(4.965,0.616)--(5.226,0.616)--(5.226,1.531)--(4.965,1.531)--cycle}
\gpfill{color=gpbgfillcolor} \gpfillpath;
\gpfill{color=gp lt color 0,gp pattern 0,pattern color=.} \gpfillpath;
\draw[gp path] (4.965,0.616)--(4.965,1.530)--(5.225,1.530)--(5.225,0.616)--cycle;
\def\gpfillpath{(6.267,0.616)--(6.528,0.616)--(6.528,2.099)--(6.267,2.099)--cycle}
\gpfill{color=gpbgfillcolor} \gpfillpath;
\gpfill{color=gp lt color 0,gp pattern 0,pattern color=.} \gpfillpath;
\draw[gp path] (6.267,0.616)--(6.267,2.098)--(6.527,2.098)--(6.527,0.616)--cycle;
\gpcolor{color=gp lt color border}
\node[gp node right] at (3.290,3.930) {GraphChi};
\def\gpfillpath{(3.474,3.853)--(4.390,3.853)--(4.390,4.007)--(3.474,4.007)--cycle}
\gpfill{color=gpbgfillcolor} \gpfillpath;
\gpfill{color=gp lt color 1,gp pattern 1,pattern color=.} \gpfillpath;
\gpcolor{color=gp lt color 1}
\gpsetlinetype{gp lt plot 1}
\draw[gp path] (3.474,3.853)--(4.390,3.853)--(4.390,4.007)--(3.474,4.007)--cycle;
\def\gpfillpath{(3.924,0.616)--(4.185,0.616)--(4.185,2.266)--(3.924,2.266)--cycle}
\gpfill{color=gpbgfillcolor} \gpfillpath;
\gpfill{color=gp lt color 1,gp pattern 1,pattern color=.} \gpfillpath;
\draw[gp path] (3.924,0.616)--(3.924,2.265)--(4.184,2.265)--(4.184,0.616)--cycle;
\def\gpfillpath{(5.225,0.616)--(5.487,0.616)--(5.487,2.496)--(5.225,2.496)--cycle}
\gpfill{color=gpbgfillcolor} \gpfillpath;
\gpfill{color=gp lt color 1,gp pattern 1,pattern color=.} \gpfillpath;
\draw[gp path] (5.225,0.616)--(5.225,2.495)--(5.486,2.495)--(5.486,0.616)--cycle;
\def\gpfillpath{(6.527,0.616)--(6.789,0.616)--(6.789,2.408)--(6.527,2.408)--cycle}
\gpfill{color=gpbgfillcolor} \gpfillpath;
\gpfill{color=gp lt color 1,gp pattern 1,pattern color=.} \gpfillpath;
\draw[gp path] (6.527,0.616)--(6.527,2.407)--(6.788,2.407)--(6.788,0.616)--cycle;
\gpcolor{color=gp lt color border}
\node[gp node right] at (6.414,4.238) {X-stream};
\def\gpfillpath{(6.598,4.161)--(7.514,4.161)--(7.514,4.315)--(6.598,4.315)--cycle}
\gpfill{color=gpbgfillcolor} \gpfillpath;
\gpfill{color=gp lt color 2,gp pattern 2,pattern color=.} \gpfillpath;
\gpcolor{color=gp lt color 2}
\gpsetlinetype{gp lt plot 2}
\draw[gp path] (6.598,4.161)--(7.514,4.161)--(7.514,4.315)--(6.598,4.315)--cycle;
\def\gpfillpath{(2.882,0.616)--(3.144,0.616)--(3.144,2.106)--(2.882,2.106)--cycle}
\gpfill{color=gpbgfillcolor} \gpfillpath;
\gpfill{color=gp lt color 2,gp pattern 2,pattern color=.} \gpfillpath;
\draw[gp path] (2.882,0.616)--(2.882,2.105)--(3.143,2.105)--(3.143,0.616)--cycle;
\def\gpfillpath{(4.184,0.616)--(4.445,0.616)--(4.445,2.152)--(4.184,2.152)--cycle}
\gpfill{color=gpbgfillcolor} \gpfillpath;
\gpfill{color=gp lt color 2,gp pattern 2,pattern color=.} \gpfillpath;
\draw[gp path] (4.184,0.616)--(4.184,2.151)--(4.444,2.151)--(4.444,0.616)--cycle;
\def\gpfillpath{(5.486,0.616)--(5.747,0.616)--(5.747,2.469)--(5.486,2.469)--cycle}
\gpfill{color=gpbgfillcolor} \gpfillpath;
\gpfill{color=gp lt color 2,gp pattern 2,pattern color=.} \gpfillpath;
\draw[gp path] (5.486,0.616)--(5.486,2.468)--(5.746,2.468)--(5.746,0.616)--cycle;
\def\gpfillpath{(6.788,0.616)--(7.049,0.616)--(7.049,2.906)--(6.788,2.906)--cycle}
\gpfill{color=gpbgfillcolor} \gpfillpath;
\gpfill{color=gp lt color 2,gp pattern 2,pattern color=.} \gpfillpath;
\draw[gp path] (6.788,0.616)--(6.788,2.905)--(7.048,2.905)--(7.048,0.616)--cycle;
\gpcolor{color=gp lt color border}
\gpsetlinetype{gp lt border}
\draw[gp path] (1.320,3.279)--(1.320,0.616)--(7.829,0.616)--(7.829,3.279)--cycle;
%% coordinates of the plot area
\gpdefrectangularnode{gp plot 1}{\pgfpoint{1.320cm}{0.616cm}}{\pgfpoint{7.829cm}{3.279cm}}
\end{tikzpicture}
%% gnuplot variables
\vspace{-15pt}
\caption{Runtime}
\label{ex:rt}
\end{subfigure}

\vspace{-15pt}
\begin{subfigure}{.5\textwidth}
\begin{tikzpicture}[gnuplot]
%% generated with GNUPLOT 4.6p4 (Lua 5.1; terminal rev. 99, script rev. 100)
%% Mon 22 Sep 2014 09:28:28 PM EDT
\path (0.000,0.000) rectangle (8.382,4.064);
\gpcolor{color=gp lt color border}
\gpsetlinetype{gp lt border}
\gpsetlinewidth{1.00}
\draw[gp path] (1.136,0.616)--(1.316,0.616);
\draw[gp path] (7.829,0.616)--(7.649,0.616);
\node[gp node right] at (0.952,0.616) { 0};
\draw[gp path] (1.136,1.012)--(1.316,1.012);
\draw[gp path] (7.829,1.012)--(7.649,1.012);
\node[gp node right] at (0.952,1.012) { 1};
\draw[gp path] (1.136,1.408)--(1.316,1.408);
\draw[gp path] (7.829,1.408)--(7.649,1.408);
\node[gp node right] at (0.952,1.408) { 2};
\draw[gp path] (1.136,1.804)--(1.316,1.804);
\draw[gp path] (7.829,1.804)--(7.649,1.804);
\node[gp node right] at (0.952,1.804) { 3};
\draw[gp path] (1.136,2.199)--(1.316,2.199);
\draw[gp path] (7.829,2.199)--(7.649,2.199);
\node[gp node right] at (0.952,2.199) { 4};
\draw[gp path] (1.136,2.595)--(1.316,2.595);
\draw[gp path] (7.829,2.595)--(7.649,2.595);
\node[gp node right] at (0.952,2.595) { 5};
\draw[gp path] (1.136,2.991)--(1.316,2.991);
\draw[gp path] (7.829,2.991)--(7.649,2.991);
\node[gp node right] at (0.952,2.991) { 6};
\draw[gp path] (1.136,3.387)--(1.316,3.387);
\draw[gp path] (7.829,3.387)--(7.649,3.387);
\node[gp node right] at (0.952,3.387) { 7};
\draw[gp path] (2.475,0.616)--(2.475,0.796);
\draw[gp path] (2.475,3.387)--(2.475,3.207);
\node[gp node center] at (2.475,0.308) {BFS};
\draw[gp path] (3.813,0.616)--(3.813,0.796);
\draw[gp path] (3.813,3.387)--(3.813,3.207);
\node[gp node center] at (3.813,0.308) {WCC};
\draw[gp path] (5.152,0.616)--(5.152,0.796);
\draw[gp path] (5.152,3.387)--(5.152,3.207);
\node[gp node center] at (5.152,0.308) {PR};
\draw[gp path] (6.490,0.616)--(6.490,0.796);
\draw[gp path] (6.490,3.387)--(6.490,3.207);
\node[gp node center] at (6.490,0.308) {TC};
\draw[gp path] (1.136,3.387)--(1.136,0.616)--(7.829,0.616)--(7.829,3.387)--cycle;
\node[gp node center,rotate=-270] at (0.246,2.001) {Memory (GB)};
\def\gpfillpath{(2.207,0.616)--(2.476,0.616)--(2.476,1.527)--(2.207,1.527)--cycle}
\gpfill{color=gpbgfillcolor} \gpfillpath;
\gpfill{color=gp lt color 0,gp pattern 0,pattern color=.} \gpfillpath;
\gpcolor{color=gp lt color 0}
\gpsetlinetype{gp lt plot 0}
\draw[gp path] (2.207,0.616)--(2.207,1.526)--(2.475,1.526)--(2.475,0.616)--cycle;
\def\gpfillpath{(3.545,0.616)--(3.814,0.616)--(3.814,1.527)--(3.545,1.527)--cycle}
\gpfill{color=gpbgfillcolor} \gpfillpath;
\gpfill{color=gp lt color 0,gp pattern 0,pattern color=.} \gpfillpath;
\draw[gp path] (3.545,0.616)--(3.545,1.526)--(3.813,1.526)--(3.813,0.616)--cycle;
\def\gpfillpath{(4.884,0.616)--(5.153,0.616)--(5.153,1.646)--(4.884,1.646)--cycle}
\gpfill{color=gpbgfillcolor} \gpfillpath;
\gpfill{color=gp lt color 0,gp pattern 0,pattern color=.} \gpfillpath;
\draw[gp path] (4.884,0.616)--(4.884,1.645)--(5.152,1.645)--(5.152,0.616)--cycle;
\def\gpfillpath{(6.223,0.616)--(6.491,0.616)--(6.491,1.884)--(6.223,1.884)--cycle}
\gpfill{color=gpbgfillcolor} \gpfillpath;
\gpfill{color=gp lt color 0,gp pattern 0,pattern color=.} \gpfillpath;
\draw[gp path] (6.223,0.616)--(6.223,1.883)--(6.490,1.883)--(6.490,0.616)--cycle;
\def\gpfillpath{(3.813,0.616)--(4.082,0.616)--(4.082,1.884)--(3.813,1.884)--cycle}
\gpfill{color=gpbgfillcolor} \gpfillpath;
\gpfill{color=gp lt color 1,gp pattern 1,pattern color=.} \gpfillpath;
\gpcolor{color=gp lt color 1}
\gpsetlinetype{gp lt plot 1}
\draw[gp path] (3.813,0.616)--(3.813,1.883)--(4.081,1.883)--(4.081,0.616)--cycle;
\def\gpfillpath{(5.152,0.616)--(5.421,0.616)--(5.421,1.132)--(5.152,1.132)--cycle}
\gpfill{color=gpbgfillcolor} \gpfillpath;
\gpfill{color=gp lt color 1,gp pattern 1,pattern color=.} \gpfillpath;
\draw[gp path] (5.152,0.616)--(5.152,1.131)--(5.420,1.131)--(5.420,0.616)--cycle;
\def\gpfillpath{(6.490,0.616)--(6.759,0.616)--(6.759,3.388)--(6.490,3.388)--cycle}
\gpfill{color=gpbgfillcolor} \gpfillpath;
\gpfill{color=gp lt color 1,gp pattern 1,pattern color=.} \gpfillpath;
\draw[gp path] (6.490,0.616)--(6.490,3.387)--(6.758,3.387)--(6.758,0.616)--cycle;
\def\gpfillpath{(2.742,0.616)--(3.011,0.616)--(3.011,1.013)--(2.742,1.013)--cycle}
\gpfill{color=gpbgfillcolor} \gpfillpath;
\gpfill{color=gp lt color 2,gp pattern 2,pattern color=.} \gpfillpath;
\gpcolor{color=gp lt color 2}
\gpsetlinetype{gp lt plot 2}
\draw[gp path] (2.742,0.616)--(2.742,1.012)--(3.010,1.012)--(3.010,0.616)--cycle;
\def\gpfillpath{(4.081,0.616)--(4.350,0.616)--(4.350,1.013)--(4.081,1.013)--cycle}
\gpfill{color=gpbgfillcolor} \gpfillpath;
\gpfill{color=gp lt color 2,gp pattern 2,pattern color=.} \gpfillpath;
\draw[gp path] (4.081,0.616)--(4.081,1.012)--(4.349,1.012)--(4.349,0.616)--cycle;
\def\gpfillpath{(5.420,0.616)--(5.688,0.616)--(5.688,1.013)--(5.420,1.013)--cycle}
\gpfill{color=gpbgfillcolor} \gpfillpath;
\gpfill{color=gp lt color 2,gp pattern 2,pattern color=.} \gpfillpath;
\draw[gp path] (5.420,0.616)--(5.420,1.012)--(5.687,1.012)--(5.687,0.616)--cycle;
\def\gpfillpath{(6.758,0.616)--(7.027,0.616)--(7.027,1.013)--(6.758,1.013)--cycle}
\gpfill{color=gpbgfillcolor} \gpfillpath;
\gpfill{color=gp lt color 2,gp pattern 2,pattern color=.} \gpfillpath;
\draw[gp path] (6.758,0.616)--(6.758,1.012)--(7.026,1.012)--(7.026,0.616)--cycle;
\gpcolor{color=gp lt color border}
\gpsetlinetype{gp lt border}
\draw[gp path] (1.136,3.387)--(1.136,0.616)--(7.829,0.616)--(7.829,3.387)--cycle;
%% coordinates of the plot area
\gpdefrectangularnode{gp plot 1}{\pgfpoint{1.136cm}{0.616cm}}{\pgfpoint{7.829cm}{3.387cm}}
\end{tikzpicture}
%% gnuplot variables
\vspace{-15pt}
\caption{Memory consumption}
\label{ex:mem}
\end{subfigure}
\caption{The runtime and memory consumption of semi-external memory FlashGraph
and external memory graph engines on the Twitter graph.}
\label{vs_ext_mem}
\end{figure}

FlashGraph outperforms GraphChi and X-Stream by one or two orders of magnitude
(Figure \ref{ex:rt}). FlashGraph only needs to access the edge lists and performs
computation on only the vertices required by the graph application. Even though FlashGraph
generates random I/O accesses, it saves both CPU and I/O by avoiding unnecessary
computation and data access. In contrast, GraphChi and X-Stream sequentially read
the entire graph dataset multiple times.
%As the authors of X-Stream point out themselves, the problem gets worse
%on a graph with a larger diameter. As we show in Table \ref{graphs}, the Twitter
%graph has a relatively small diameter, so we anticipate both X-Stream
%and GraphChi's performance will degrade with the other graphs.

%\footnotetext[1]{We are unable to reproduce the BFS result shown in
%the original paper \cite{xstream}}

Although FlashGraph uses its semi-external memory mode, it consumes
a reasonable amount of memory when compared with GraphChi and X-Stream
(Figure \ref{ex:mem}). In some applications, FlashGraph even has smaller
memory footprint than GraphChi. FlashGraph's small memory footprint
allows it to run on regular desktop computers, comfortably processing
billion-edge graphs.
%The performance of triangle counting in GraphChi highly depends on
%the memory size available for computation. It requires a considerable
%amount of memory to achieve reasonable performance.

\subsection{Scale to billion-node graphs}
We further evaluate the performance of FlashGraph on the billion-scale page
graph in Table \ref{graphs}. FlashGraph uses a page cache of 4GB for all
applications. To the best of our knowledge, the page graph
is the largest graph used for evaluating a graph processing engine to date.
The closest one is the random graph used by Pregel \cite{pregel}, which has
a billion vertices and $127$ billion edges. Pregel
processed it on $300$ multicore machines. In contrast, we process the page
graph on a single multicore machine.

FlashGraph can perform all of our applications within a reasonable amount of
time and with relatively small memory footprint (Table \ref{page_graph}).
For example, FlashGraph achieves good performance in BFS on this billion-node
graph. It takes less than
five minutes with a cache size of 4GB; i.e., FlashGraph traverses
nearly seven million vertices per second on the page graph, which is much higher than
the maximal random I/O performance ($900,000$ IOPS) provided by the SSD array.
In contrast, Pregel \cite{pregel} used $300$ multicore machines to run the shortest
path algorithm on their
largest random graph and took a little over ten minutes. More recently, Trinity
\cite{trinity} took over ten minutes to perform BFS on a graph of
one billion vertices and $13$ billion edges on $14$ $12$-core machines.

Our solution allows us to process a graph one order of magnitude larger
than the page graph on a single commodity machine with half a terabyte of RAM.
The maximal graph size that can be processed by FlashGraph is limited by
the capacity of RAM and SSDs. Our current hardware configuration allows
us to attach 24 1TB SSDs to a machine, which can store a graph with over one trillion
edges. Furthermore, the small memory footprint suggests that FlashGraph is
able to process a graph with tens of billions of vertices.

FlashGraph results in a more economical solution to process a massive graph.
In contrast, it is much more expensive to build a cluster or a supercomputer
to process a graph of the same scale. For example, it requires $48$ machines
with 512GB RAM each to achieve 24TB aggregate RAM capacity, so the cost of
building such a cluster is at least $24-48$ times higher than our solution.
In addition, FlashGraph minimizes SSD wearout and the only write required
by FlashGraph is to load a new graph to SSDs for processing. Therefore, we
can further reduce the hardware cost, by using consumer SSDs instead of
enterprise SSDs to store graphs, as well as reducing the maintenance cost.
%makes massive graph analysis more accessible to users, who traditionally would
%require a large cluster to perform graph computations at this scale.

%As suggested by the results in Table \ref{page_graph}, a larger page cache
%does not always improve performance for some graph algorithms significantly.
%Breadth-first search gets the largest performance improvement of $41$\%,
%while finding connected components is not improved by a larger cache
%size at all. Triangle counting and scan statistics can generate
%relatively good cache hit rates because the vertices in the page graph are
%relatively well clustered by their respective domains.

\begin{table}
\begin{center}
\footnotesize
\begin{tabular}{|c|c|c|c|c|}
\hline
Algorithm & Runtime (sec) & Init time (sec) & Memory (GB) \\
\hline
BFS & $298$ & $30$ & $22$ \\
\hline
BC & $595$ & $33$ & $81$ \\
\hline
TC & $7818$ & $31$ & $55$ \\
\hline
WCC & $461$ & $32$ & $47$ \\
\hline
PR & $2041$ & $33$ & $46$ \\
\hline
SS & $375$ & $58$ & $83$ \\
\hline
\end{tabular}
\normalsize
\end{center}
\caption{The runtime and memory consumption of FlashGraph
on the page graph using a $4$GB cache size.}
\label{page_graph}
\end{table}

\subsection{The impact of optimizations}
In this section, we perform experiments to justify some of our design decisions
that are critical to achieve performance for FlashGraph in semi-external memory.

\subsubsection{Preserve sequential I/O} \label{sec_preserve_seq_io}
We demonstrate the importance of taking advantage of sequential I/O access in
graph applications, using BFS and weakly connected
components. We start with vertex execution performed in random order,
and then sequentially order vertex execution by vertex ID. 
Finally, we show the performance difference
between merging I/O requests in SAFS vs.~FlashGraph. All experiments are run on
the subdomain web graph.

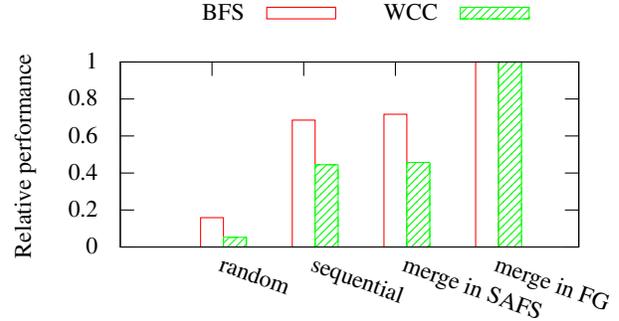
\begin{figure}[tb]
	\begin{center}
		\small
		\vspace{-15pt}
		\begin{tikzpicture}[gnuplot]
%% generated with GNUPLOT 4.6p4 (Lua 5.1; terminal rev. 99, script rev. 100)
%% Sat 03 Jan 2015 11:42:02 PM EST
\path (0.000,0.000) rectangle (8.382,4.572);
\gpcolor{color=gp lt color border}
\gpsetlinetype{gp lt border}
\gpsetlinewidth{1.00}
\draw[gp path] (1.504,1.126)--(1.684,1.126);
\draw[gp path] (7.599,1.126)--(7.419,1.126);
\node[gp node right] at (1.320,1.126) { 0};
\draw[gp path] (1.504,1.618)--(1.684,1.618);
\draw[gp path] (7.599,1.618)--(7.419,1.618);
\node[gp node right] at (1.320,1.618) { 0.2};
\draw[gp path] (1.504,2.110)--(1.684,2.110);
\draw[gp path] (7.599,2.110)--(7.419,2.110);
\node[gp node right] at (1.320,2.110) { 0.4};
\draw[gp path] (1.504,2.603)--(1.684,2.603);
\draw[gp path] (7.599,2.603)--(7.419,2.603);
\node[gp node right] at (1.320,2.603) { 0.6};
\draw[gp path] (1.504,3.095)--(1.684,3.095);
\draw[gp path] (7.599,3.095)--(7.419,3.095);
\node[gp node right] at (1.320,3.095) { 0.8};
\draw[gp path] (1.504,3.587)--(1.684,3.587);
\draw[gp path] (7.599,3.587)--(7.419,3.587);
\node[gp node right] at (1.320,3.587) { 1};
\draw[gp path] (2.723,1.126)--(2.723,1.306);
\draw[gp path] (2.723,3.587)--(2.723,3.407);
\node[gp node left,rotate=-20] at (2.723,0.942) {random};
\draw[gp path] (3.942,1.126)--(3.942,1.306);
\draw[gp path] (3.942,3.587)--(3.942,3.407);
\node[gp node left,rotate=-20] at (3.942,0.942) {sequential};
\draw[gp path] (5.161,1.126)--(5.161,1.306);
\draw[gp path] (5.161,3.587)--(5.161,3.407);
\node[gp node left,rotate=-20] at (5.161,0.942) {merge in SAFS};
\draw[gp path] (6.380,1.126)--(6.380,1.306);
\draw[gp path] (6.380,3.587)--(6.380,3.407);
\node[gp node left,rotate=-20] at (6.380,0.942) {merge in FG};
\draw[gp path] (1.504,3.587)--(1.504,1.126)--(7.599,1.126)--(7.599,3.587)--cycle;
\node[gp node center,rotate=-270] at (0.246,2.356) {Relative performance};
\node[gp node right] at (3.267,4.238) {BFS};
\def\gpfillpath{(3.451,4.161)--(4.367,4.161)--(4.367,4.315)--(3.451,4.315)--cycle}
\gpfill{color=gpbgfillcolor} \gpfillpath;
\gpfill{color=gp lt color 0,gp pattern 0,pattern color=.} \gpfillpath;
\gpcolor{color=gp lt color 0}
\gpsetlinetype{gp lt plot 0}
\draw[gp path] (3.451,4.161)--(4.367,4.161)--(4.367,4.315)--(3.451,4.315)--cycle;
\def\gpfillpath{(2.571,1.126)--(2.876,1.126)--(2.876,1.517)--(2.571,1.517)--cycle}
\gpfill{color=gpbgfillcolor} \gpfillpath;
\gpfill{color=gp lt color 0,gp pattern 0,pattern color=.} \gpfillpath;
\draw[gp path] (2.571,1.126)--(2.571,1.516)--(2.875,1.516)--(2.875,1.126)--cycle;
\def\gpfillpath{(3.790,1.126)--(4.095,1.126)--(4.095,2.815)--(3.790,2.815)--cycle}
\gpfill{color=gpbgfillcolor} \gpfillpath;
\gpfill{color=gp lt color 0,gp pattern 0,pattern color=.} \gpfillpath;
\draw[gp path] (3.790,1.126)--(3.790,2.814)--(4.094,2.814)--(4.094,1.126)--cycle;
\def\gpfillpath{(5.009,1.126)--(5.314,1.126)--(5.314,2.893)--(5.009,2.893)--cycle}
\gpfill{color=gpbgfillcolor} \gpfillpath;
\gpfill{color=gp lt color 0,gp pattern 0,pattern color=.} \gpfillpath;
\draw[gp path] (5.009,1.126)--(5.009,2.892)--(5.313,2.892)--(5.313,1.126)--cycle;
\def\gpfillpath{(6.228,1.126)--(6.533,1.126)--(6.533,3.588)--(6.228,3.588)--cycle}
\gpfill{color=gpbgfillcolor} \gpfillpath;
\gpfill{color=gp lt color 0,gp pattern 0,pattern color=.} \gpfillpath;
\draw[gp path] (6.228,1.126)--(6.228,3.587)--(6.532,3.587)--(6.532,1.126)--cycle;
\gpcolor{color=gp lt color border}
\node[gp node right] at (5.839,4.238) {    WCC};
\def\gpfillpath{(6.023,4.161)--(6.939,4.161)--(6.939,4.315)--(6.023,4.315)--cycle}
\gpfill{color=gpbgfillcolor} \gpfillpath;
\gpfill{color=gp lt color 1,gp pattern 1,pattern color=.} \gpfillpath;
\gpcolor{color=gp lt color 1}
\gpsetlinetype{gp lt plot 1}
\draw[gp path] (6.023,4.161)--(6.939,4.161)--(6.939,4.315)--(6.023,4.315)--cycle;
\def\gpfillpath{(2.875,1.126)--(3.181,1.126)--(3.181,1.255)--(2.875,1.255)--cycle}
\gpfill{color=gpbgfillcolor} \gpfillpath;
\gpfill{color=gp lt color 1,gp pattern 1,pattern color=.} \gpfillpath;
\draw[gp path] (2.875,1.126)--(2.875,1.254)--(3.180,1.254)--(3.180,1.126)--cycle;
\def\gpfillpath{(4.094,1.126)--(4.400,1.126)--(4.400,2.220)--(4.094,2.220)--cycle}
\gpfill{color=gpbgfillcolor} \gpfillpath;
\gpfill{color=gp lt color 1,gp pattern 1,pattern color=.} \gpfillpath;
\draw[gp path] (4.094,1.126)--(4.094,2.219)--(4.399,2.219)--(4.399,1.126)--cycle;
\def\gpfillpath{(5.313,1.126)--(5.619,1.126)--(5.619,2.251)--(5.313,2.251)--cycle}
\gpfill{color=gpbgfillcolor} \gpfillpath;
\gpfill{color=gp lt color 1,gp pattern 1,pattern color=.} \gpfillpath;
\draw[gp path] (5.313,1.126)--(5.313,2.250)--(5.618,2.250)--(5.618,1.126)--cycle;
\def\gpfillpath{(6.532,1.126)--(6.838,1.126)--(6.838,3.588)--(6.532,3.588)--cycle}
\gpfill{color=gpbgfillcolor} \gpfillpath;
\gpfill{color=gp lt color 1,gp pattern 1,pattern color=.} \gpfillpath;
\draw[gp path] (6.532,1.126)--(6.532,3.587)--(6.837,3.587)--(6.837,1.126)--cycle;
\gpcolor{color=gp lt color border}
\gpsetlinetype{gp lt border}
\draw[gp path] (1.504,3.587)--(1.504,1.126)--(7.599,1.126)--(7.599,3.587)--cycle;
%% coordinates of the plot area
\gpdefrectangularnode{gp plot 1}{\pgfpoint{1.504cm}{1.126cm}}{\pgfpoint{7.599cm}{3.587cm}}
\end{tikzpicture}
%% gnuplot variables
		\vspace{-15pt}
		\caption{The impact of preserving sequential I/O access in graph
		applications. All performance is relative to that of
		the implementation of merging I/O requests in FlashGraph.}
		\label{preserve_seq}
	\end{center}
\end{figure}

The huge gap (Figure \ref{preserve_seq}) between random execution and sequential
execution suggests that there exists a degree of sequential I/O in both
applications, as described in Section \ref{vertex_access}.
% Can you live without these next four lines??
%WCC requires access to all edge lists in the first iteration,
%so it is not surprising that this application has sequential I/O access.
%BFS visits many vertices in some iterations and issues many I/O
%requests in these iterations to access edge lists, which are likely stored in
%adjacent pages on SSDs. 
If FlashGraph did not take advantage of these sequential
I/O accesses, it would suffer substantial performance degradation. Therefore,
the first priority of the vertex scheduler in FlashGraph
is to schedule vertex execution to generate sequential I/O. Consequently,
FlashGraph's vertex scheduler is highly constrained by I/O ordering requirements and
is not able to schedule vertex execution freely like Galois \cite{galois}.

Figure \ref{preserve_seq} also shows that I/O accesses generated by a graph
algorithm are well merged
in FlashGraph as opposed to the filesystem level or the block subsystem
level. Although SAFS, the Linux filesystem and the Linux block subsystem are
capable of merging I/O requests, they require more CPU computation to merge
I/O requests and do not have a global view for merging
I/O requests. Consequently, it is much more light-weight and effective to
merge I/O requests in FlashGraph. By doing so,
we achieve 40\% speedup for BFS and more than 100\% speedup for WCC.

\subsubsection{The impact of the page size}
In this section, we investigate the impact of the page size in SAFS. A page is
the smallest I/O block that FlashGraph can access from SSDs. The experiments
are run on the subdomain web graph.

\begin{figure}[tb]
	\begin{center}
		\small
		\vspace{-15pt}
		\begin{tikzpicture}[gnuplot]
%% generated with GNUPLOT 4.6p4 (Lua 5.1; terminal rev. 99, script rev. 100)
%% Tue 20 Jan 2015 01:45:51 PM EST
\path (0.000,0.000) rectangle (8.382,4.318);
\gpcolor{color=gp lt color border}
\gpsetlinetype{gp lt border}
\gpsetlinewidth{1.00}
\draw[gp path] (1.504,0.985)--(1.684,0.985);
\draw[gp path] (7.829,0.985)--(7.649,0.985);
\node[gp node right] at (1.320,0.985) { 0};
\draw[gp path] (1.504,1.376)--(1.684,1.376);
\draw[gp path] (7.829,1.376)--(7.649,1.376);
\node[gp node right] at (1.320,1.376) { 0.2};
\draw[gp path] (1.504,1.768)--(1.684,1.768);
\draw[gp path] (7.829,1.768)--(7.649,1.768);
\node[gp node right] at (1.320,1.768) { 0.4};
\draw[gp path] (1.504,2.159)--(1.684,2.159);
\draw[gp path] (7.829,2.159)--(7.649,2.159);
\node[gp node right] at (1.320,2.159) { 0.6};
\draw[gp path] (1.504,2.550)--(1.684,2.550);
\draw[gp path] (7.829,2.550)--(7.649,2.550);
\node[gp node right] at (1.320,2.550) { 0.8};
\draw[gp path] (1.504,2.942)--(1.684,2.942);
\draw[gp path] (7.829,2.942)--(7.649,2.942);
\node[gp node right] at (1.320,2.942) { 1};
\draw[gp path] (1.504,3.333)--(1.684,3.333);
\draw[gp path] (7.829,3.333)--(7.649,3.333);
\node[gp node right] at (1.320,3.333) { 1.2};
\draw[gp path] (1.504,0.985)--(1.504,1.165);
\draw[gp path] (1.504,3.333)--(1.504,3.153);
\node[gp node center] at (1.504,0.677) {1};
\draw[gp path] (2.137,0.985)--(2.137,1.165);
\draw[gp path] (2.137,3.333)--(2.137,3.153);
\node[gp node center] at (2.137,0.677) {2};
\draw[gp path] (2.769,0.985)--(2.769,1.165);
\draw[gp path] (2.769,3.333)--(2.769,3.153);
\node[gp node center] at (2.769,0.677) {4};
\draw[gp path] (3.402,0.985)--(3.402,1.165);
\draw[gp path] (3.402,3.333)--(3.402,3.153);
\node[gp node center] at (3.402,0.677) {8};
\draw[gp path] (4.034,0.985)--(4.034,1.165);
\draw[gp path] (4.034,3.333)--(4.034,3.153);
\node[gp node center] at (4.034,0.677) {16};
\draw[gp path] (4.667,0.985)--(4.667,1.165);
\draw[gp path] (4.667,3.333)--(4.667,3.153);
\node[gp node center] at (4.667,0.677) {32};
\draw[gp path] (5.299,0.985)--(5.299,1.165);
\draw[gp path] (5.299,3.333)--(5.299,3.153);
\node[gp node center] at (5.299,0.677) {64};
\draw[gp path] (5.932,0.985)--(5.932,1.165);
\draw[gp path] (5.932,3.333)--(5.932,3.153);
\node[gp node center] at (5.932,0.677) {128};
\draw[gp path] (6.564,0.985)--(6.564,1.165);
\draw[gp path] (6.564,3.333)--(6.564,3.153);
\node[gp node center] at (6.564,0.677) {256};
\draw[gp path] (7.197,0.985)--(7.197,1.165);
\draw[gp path] (7.197,3.333)--(7.197,3.153);
\node[gp node center] at (7.197,0.677) {512};
\draw[gp path] (7.829,0.985)--(7.829,1.165);
\draw[gp path] (7.829,3.333)--(7.829,3.153);
\node[gp node center] at (7.829,0.677) {1024};
\draw[gp path] (1.504,3.333)--(1.504,0.985)--(7.829,0.985)--(7.829,3.333)--cycle;
\node[gp node center,rotate=-270] at (0.246,2.159) {Relative performance};
\node[gp node center] at (4.666,0.215) {Page size (KB)};
\node[gp node right] at (2.464,3.984) {BFS};
\gpcolor{color=gp lt color 0}
\gpsetlinetype{gp lt plot 0}
\draw[gp path] (2.648,3.984)--(3.564,3.984);
\draw[gp path] (1.504,2.237)--(2.137,2.707)--(2.769,2.942)--(3.402,2.883)--(4.034,2.492)%
  --(4.667,2.296)--(5.299,2.159)--(5.932,2.022)--(6.564,1.924)--(7.197,1.787)--(7.829,1.650);
\gpsetpointsize{4.00}
\gppoint{gp mark 1}{(1.504,2.237)}
\gppoint{gp mark 1}{(2.137,2.707)}
\gppoint{gp mark 1}{(2.769,2.942)}
\gppoint{gp mark 1}{(3.402,2.883)}
\gppoint{gp mark 1}{(4.034,2.492)}
\gppoint{gp mark 1}{(4.667,2.296)}
\gppoint{gp mark 1}{(5.299,2.159)}
\gppoint{gp mark 1}{(5.932,2.022)}
\gppoint{gp mark 1}{(6.564,1.924)}
\gppoint{gp mark 1}{(7.197,1.787)}
\gppoint{gp mark 1}{(7.829,1.650)}
\gppoint{gp mark 1}{(3.106,3.984)}
\gpcolor{color=gp lt color border}
\node[gp node right] at (4.300,3.984) {WCC};
\gpcolor{color=gp lt color 1}
\gpsetlinetype{gp lt plot 1}
\draw[gp path] (4.484,3.984)--(5.400,3.984);
\draw[gp path] (1.504,2.433)--(2.137,2.687)--(2.769,2.942)--(3.402,2.903)--(4.034,2.805)%
  --(4.667,2.726)--(5.299,2.746)--(5.932,2.609)--(6.564,2.511)--(7.197,2.413)--(7.829,2.316);
\gppoint{gp mark 2}{(1.504,2.433)}
\gppoint{gp mark 2}{(2.137,2.687)}
\gppoint{gp mark 2}{(2.769,2.942)}
\gppoint{gp mark 2}{(3.402,2.903)}
\gppoint{gp mark 2}{(4.034,2.805)}
\gppoint{gp mark 2}{(4.667,2.726)}
\gppoint{gp mark 2}{(5.299,2.746)}
\gppoint{gp mark 2}{(5.932,2.609)}
\gppoint{gp mark 2}{(6.564,2.511)}
\gppoint{gp mark 2}{(7.197,2.413)}
\gppoint{gp mark 2}{(7.829,2.316)}
\gppoint{gp mark 2}{(4.942,3.984)}
\gpcolor{color=gp lt color border}
\node[gp node right] at (6.136,3.984) {TC};
\gpcolor{color=gp lt color 2}
\gpsetlinetype{gp lt plot 2}
\draw[gp path] (6.320,3.984)--(7.236,3.984);
\draw[gp path] (1.504,2.824)--(2.137,2.981)--(2.769,2.942)--(3.402,2.629)--(4.034,1.963)%
  --(4.667,1.729)--(5.299,1.533)--(5.932,1.337)--(6.564,1.200)--(7.197,1.102)--(7.829,1.044);
\gppoint{gp mark 3}{(1.504,2.824)}
\gppoint{gp mark 3}{(2.137,2.981)}
\gppoint{gp mark 3}{(2.769,2.942)}
\gppoint{gp mark 3}{(3.402,2.629)}
\gppoint{gp mark 3}{(4.034,1.963)}
\gppoint{gp mark 3}{(4.667,1.729)}
\gppoint{gp mark 3}{(5.299,1.533)}
\gppoint{gp mark 3}{(5.932,1.337)}
\gppoint{gp mark 3}{(6.564,1.200)}
\gppoint{gp mark 3}{(7.197,1.102)}
\gppoint{gp mark 3}{(7.829,1.044)}
\gppoint{gp mark 3}{(6.778,3.984)}
\gpcolor{color=gp lt color border}
\gpsetlinetype{gp lt border}
\draw[gp path] (1.504,3.333)--(1.504,0.985)--(7.829,0.985)--(7.829,3.333)--cycle;
%% coordinates of the plot area
\gpdefrectangularnode{gp plot 1}{\pgfpoint{1.504cm}{0.985cm}}{\pgfpoint{7.829cm}{3.333cm}}
\end{tikzpicture}
%% gnuplot variables
		\vspace{-15pt}
		\caption{The impact of the page size in FlashGraph. All performance is
		relative to that of the implementation with 4KB page size.}
		\label{block_size}
	\end{center}
\end{figure}
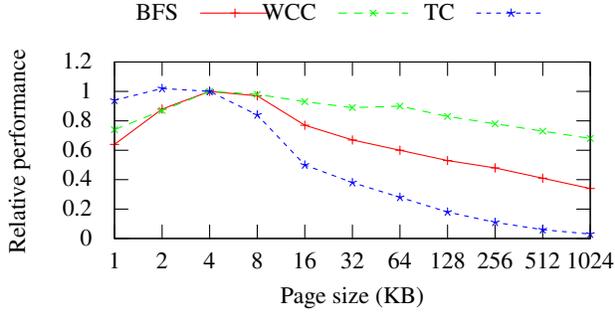

% Minimum block size is different from block size. We still want large block size.
% but we want minimum block size to be as small as possible so that the actual
% block size can vary from a page size to a very large size.
Figure \ref{block_size} shows that FlashGraph should use 4KB as
the SAFS page size. SSDs store and access data at the granularity
of 4KB flash pages, so using an SAFS page smaller than 4KB does not increase
the I/O rate of SSDs much.
A larger SAFS page size brings in more unnecessary data and
wastes I/O bandwidth, which leads to performance degradation.
When we increase the SAFS page size from 4KB to 1MB, the performance of
BFS and triangle counting (TC) decreases to a small fraction of their
maximal performance. Even WCC,
whose I/O access is more sequential, performs better with 4KB pages
because WCC also needs to selectively access edge lists in all
iterations but the first. This result suggests that TurboGraph
\cite{turbograph}, which uses a block size of multiple megabytes, may perform
general graph analysis suboptimally. It also suggests that when using
4KB pages,
selectively accessing edge lists and merging I/O enables FlashGraph
to adapt to different I/O access patterns. % resulting in near optimal performance for
%a variety of applications.

\subsection{The impact of page cache size}
We investigate the effect of the SAFS page cache size on the performance
of FlashGraph. We vary the cache size from $1$ GB to $32$GB,
which is sufficiently large to accommodate the twitter graph and
the subdomain web graph.
%We show the results on the subdomain graph in Figure \ref{cache}.
We omit Twitter graph results as they mirror subdomain graph results.

\begin{figure}[t]
\centering
\footnotesize
\vspace{-15pt}
%\begin{subfigure}{.5\textwidth}
\begin{tikzpicture}[gnuplot]
%% generated with GNUPLOT 4.6p4 (Lua 5.1; terminal rev. 99, script rev. 100)
%% Tue 23 Sep 2014 06:07:44 PM EDT
\path (0.000,0.000) rectangle (8.382,4.572);
\gpcolor{color=gp lt color border}
\gpsetlinetype{gp lt border}
\gpsetlinewidth{1.00}
\draw[gp path] (1.504,0.985)--(1.684,0.985);
\draw[gp path] (7.829,0.985)--(7.649,0.985);
\node[gp node right] at (1.320,0.985) { 0};
\draw[gp path] (1.504,1.444)--(1.684,1.444);
\draw[gp path] (7.829,1.444)--(7.649,1.444);
\node[gp node right] at (1.320,1.444) { 0.2};
\draw[gp path] (1.504,1.903)--(1.684,1.903);
\draw[gp path] (7.829,1.903)--(7.649,1.903);
\node[gp node right] at (1.320,1.903) { 0.4};
\draw[gp path] (1.504,2.361)--(1.684,2.361);
\draw[gp path] (7.829,2.361)--(7.649,2.361);
\node[gp node right] at (1.320,2.361) { 0.6};
\draw[gp path] (1.504,2.820)--(1.684,2.820);
\draw[gp path] (7.829,2.820)--(7.649,2.820);
\node[gp node right] at (1.320,2.820) { 0.8};
\draw[gp path] (1.504,3.279)--(1.684,3.279);
\draw[gp path] (7.829,3.279)--(7.649,3.279);
\node[gp node right] at (1.320,3.279) { 1};
\draw[gp path] (1.504,0.985)--(1.504,1.165);
\draw[gp path] (1.504,3.279)--(1.504,3.099);
\node[gp node center] at (1.504,0.677) {1};
\draw[gp path] (2.769,0.985)--(2.769,1.165);
\draw[gp path] (2.769,3.279)--(2.769,3.099);
\node[gp node center] at (2.769,0.677) {2};
\draw[gp path] (4.034,0.985)--(4.034,1.165);
\draw[gp path] (4.034,3.279)--(4.034,3.099);
\node[gp node center] at (4.034,0.677) {4};
\draw[gp path] (5.299,0.985)--(5.299,1.165);
\draw[gp path] (5.299,3.279)--(5.299,3.099);
\node[gp node center] at (5.299,0.677) {8};
\draw[gp path] (6.564,0.985)--(6.564,1.165);
\draw[gp path] (6.564,3.279)--(6.564,3.099);
\node[gp node center] at (6.564,0.677) {16};
\draw[gp path] (7.829,0.985)--(7.829,1.165);
\draw[gp path] (7.829,3.279)--(7.829,3.099);
\node[gp node center] at (7.829,0.677) {32};
\draw[gp path] (1.504,3.279)--(1.504,0.985)--(7.829,0.985)--(7.829,3.279)--cycle;
\node[gp node center,rotate=-270] at (0.246,2.132) {Relative performance};
\node[gp node center] at (4.666,0.215) {Cache size (GB)};
\node[gp node right] at (2.464,4.238) {BFS};
\gpcolor{color=gp lt color 0}
\gpsetlinetype{gp lt plot 0}
\draw[gp path] (2.648,4.238)--(3.564,4.238);
\draw[gp path] (1.504,2.545)--(2.769,2.660)--(4.034,2.866)--(5.232,3.279);
\draw[gp path] (5.932,3.279)--(6.564,3.256)--(7.829,3.279);
\gpsetpointsize{4.00}
\gppoint{gp mark 1}{(1.504,2.545)}
\gppoint{gp mark 1}{(2.769,2.660)}
\gppoint{gp mark 1}{(4.034,2.866)}
\gppoint{gp mark 1}{(6.564,3.256)}
\gppoint{gp mark 1}{(7.829,3.279)}
\gppoint{gp mark 1}{(3.106,4.238)}
\gpcolor{color=gp lt color border}
\node[gp node right] at (2.464,3.930) {BC};
\gpcolor{color=gp lt color 1}
\gpsetlinetype{gp lt plot 1}
\draw[gp path] (2.648,3.930)--(3.564,3.930);
\draw[gp path] (1.504,3.004)--(2.769,3.050)--(4.034,3.164)--(5.299,3.256)--(6.564,3.279)%
  --(7.829,3.279);
\gppoint{gp mark 2}{(1.504,3.004)}
\gppoint{gp mark 2}{(2.769,3.050)}
\gppoint{gp mark 2}{(4.034,3.164)}
\gppoint{gp mark 2}{(5.299,3.256)}
\gppoint{gp mark 2}{(6.564,3.279)}
\gppoint{gp mark 2}{(7.829,3.279)}
\gppoint{gp mark 2}{(3.106,3.930)}
\gpcolor{color=gp lt color border}
\node[gp node right] at (4.300,4.238) {WCC};
\gpcolor{color=gp lt color 2}
\gpsetlinetype{gp lt plot 2}
\draw[gp path] (4.484,4.238)--(5.400,4.238);
\draw[gp path] (1.504,3.095)--(2.769,3.073)--(4.034,3.073)--(5.299,3.095)--(6.564,3.233)%
  --(7.829,3.279);
\gppoint{gp mark 3}{(1.504,3.095)}
\gppoint{gp mark 3}{(2.769,3.073)}
\gppoint{gp mark 3}{(4.034,3.073)}
\gppoint{gp mark 3}{(5.299,3.095)}
\gppoint{gp mark 3}{(6.564,3.233)}
\gppoint{gp mark 3}{(7.829,3.279)}
\gppoint{gp mark 3}{(4.942,4.238)}
\gpcolor{color=gp lt color border}
\node[gp node right] at (4.300,3.930) {PR};
\gpcolor{color=gp lt color 3}
\gpsetlinetype{gp lt plot 3}
\draw[gp path] (4.484,3.930)--(5.400,3.930);
\draw[gp path] (1.504,2.706)--(2.769,2.706)--(4.034,2.728)--(5.299,3.118)--(6.564,3.279)%
  --(7.829,3.279);
\gppoint{gp mark 4}{(1.504,2.706)}
\gppoint{gp mark 4}{(2.769,2.706)}
\gppoint{gp mark 4}{(4.034,2.728)}
\gppoint{gp mark 4}{(5.299,3.118)}
\gppoint{gp mark 4}{(6.564,3.279)}
\gppoint{gp mark 4}{(7.829,3.279)}
\gppoint{gp mark 4}{(4.942,3.930)}
\gpcolor{color=gp lt color border}
\node[gp node right] at (6.136,4.238) {SS};
\gpcolor{color=gp lt color 4}
\gpsetlinetype{gp lt plot 4}
\draw[gp path] (6.320,4.238)--(7.236,4.238);
\draw[gp path] (1.504,2.591)--(2.769,2.660)--(4.034,2.660)--(5.299,3.095)--(6.564,3.118)%
  --(7.829,3.279);
\gppoint{gp mark 5}{(1.504,2.591)}
\gppoint{gp mark 5}{(2.769,2.660)}
\gppoint{gp mark 5}{(4.034,2.660)}
\gppoint{gp mark 5}{(5.299,3.095)}
\gppoint{gp mark 5}{(6.564,3.118)}
\gppoint{gp mark 5}{(7.829,3.279)}
\gppoint{gp mark 5}{(6.778,4.238)}
\gpcolor{color=gp lt color border}
\node[gp node right] at (6.136,3.930) {TC};
\gpcolor{rgb color={1.000,0.647,0.000}}
\gpsetlinetype{gp lt plot 5}
\draw[gp path] (6.320,3.930)--(7.236,3.930);
\draw[gp path] (1.504,2.499)--(2.769,2.935)--(4.034,3.118)--(5.299,3.164)--(6.564,3.210)%
  --(7.829,3.279);
\gppoint{gp mark 6}{(1.504,2.499)}
\gppoint{gp mark 6}{(2.769,2.935)}
\gppoint{gp mark 6}{(4.034,3.118)}
\gppoint{gp mark 6}{(5.299,3.164)}
\gppoint{gp mark 6}{(6.564,3.210)}
\gppoint{gp mark 6}{(7.829,3.279)}
\gppoint{gp mark 6}{(6.778,3.930)}
\gpcolor{color=gp lt color border}
\gpsetlinetype{gp lt border}
\draw[gp path] (1.504,3.279)--(1.504,0.985)--(7.829,0.985)--(7.829,3.279)--cycle;
%% coordinates of the plot area
\gpdefrectangularnode{gp plot 1}{\pgfpoint{1.504cm}{0.985cm}}{\pgfpoint{7.829cm}{3.279cm}}
\end{tikzpicture}
%% gnuplot variables
\vspace{-15pt}
%\caption{The runtime of each application is normalized to that
%of the application with 32GB cache}
%\label{cache:perf}
%\end{subfigure}

%\begin{subfigure}{.5\textwidth}
%\include{diff.cache.hit.rate}
%\vspace{-15pt}
%\caption{The cache hit rate.}
%\label{cache:rate}
%\end{subfigure}
\caption{The impact of cache size in FlashGraph.}
\label{cache}
\end{figure}
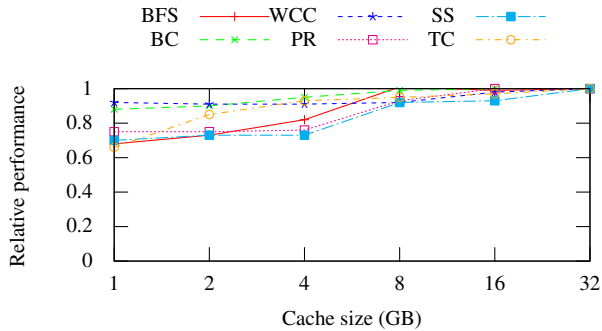

FlashGraph performs well even with a small page cache (Figure \ref{cache}).
With a $1$GB page cache, all applications realize at least 65\%
of their performance with a $32$GB page cache, and WCC and betweenness centrality
even achieve around 90\% of the performance with a $32$GB page cache.
Although PageRank has
a similar I/O access pattern to WCC, it converges more slowly than WCC,
so a large cache has more impact on PageRank. %BFS and triangle counting
%are relatively more sensitive to the cache size than other applications
%because these two applications generate more small random reads from SSDs.
By varying the page cache size, we show FlashGraph can smoothly 
transition from a semi-external memory graph engine to an in-memory graph engine.

%FlashGraph has much smaller overhead for transition between iterations,
%which is more noticeable in a large-diameter graph.
%Therefore, FlashGraph with a cache size of $1$ GB can outperform PowerGraph
%in breadth-first search by a factor of five on the subdomain web graph.
%Since PowerGraph uses a synchronous execution engine, its implementations
%for finding weakly connected components and computing PageRank converge
%at a slower rate. Furthermore, PowerGraph cannot perform effective pruning
%when computing scan statistics so it needs to perform actual computation
%on many more vertices than necessary FlashGraph.

%FlashGraph has much smaller memory footprint than PowerGraph
%(Figure \ref{memory}). FlashGraph only needs to maintain
%vertex state in memory and access data on SSDs via the page
%cache. Furthermore, FlashGraph's programming interface enables triangle
%counting and scan statistics to only need to maintain
%complex data structures when vertices are in the running state. In contrast,
%PowerGraph maintains all data in memory. It also requires much more memory to perform triangle
%counting and scan statistics because these two applications require every vertex
%to maintain much more complex data structures in PowerGraph.

%\begin{figure}[tb]
%	\begin{center}
%		\footnotesize
%		\vspace{-15pt}
%		\include{memory.vs.powergraph}
%		\vspace{-15pt}
%		\caption{The memory consumption of the applications in FlashGraph (FG) with
%		the page cache configuration of 1GB and PowerGraph (PG).}
%		\label{memory}
%	\end{center}
%\end{figure}

\section{Conclusions}
We present the semi-external memory graph engine called FlashGraph that
closely integrates with an SSD filesystem to achieve maximal performance.
It uses an asynchronous user-task I/O interface
to reduce overhead associated with accessing data in the filesystem and overlap
computation with I/O. %The graph engine assigns vertices of a graph
%to different partitions to maximize parallelism and localize
%computation to the local memory in a non-uniform memory architecture.
FlashGraph selectively accesses edge lists required by a graph
algorithm from SSDs to reduce data access;
it conservatively merges I/O requests to increase I/O throughput and
reduce CPU consumption;
it further schedules the order of processing vertices to help merge
I/O requests and maximize the page cache hit rate.
All of these designs maximize
performance for applications with different I/O access patterns.
We demonstrate that a semi-external memory graph engine
can achieve performance comparable to in-memory graph engines.

We observe that in many graph applications a large SSD array is capable
of delivering enough I/Os to saturate the CPU. This suggests the importance
of optimizing for CPU and RAM in such an I/O system. It also suggests
that SSDs have been sufficiently fast to be an important extension for RAM
when we build a machine for large-scale graph analysis applications.

FlashGraph provides a concise and flexible programming interface to express
a wide variety of graph algorithms and their optimizations. Users express
graph algorithms in FlashGraph from the perspective of vertices. Vertices
can interact with any other vertices in the graph by sending messages, which
localizes user computation to the local memory and avoids concurrent data
access to algorithmic vertex state.

Unlike other external-memory graph engines such as GraphChi and X-stream,
FlashGraph supports selective access to edge lists. We demonstrate
that streaming the entire graph to reduce random I/O leads to a suboptimal
solution for high-speed SSDs. Reading and computing on data only required
by graph applications saves computation and increases the I/O access
rate to the SSDs.

We further demonstrate that FlashGraph is able to process graphs with billions
of vertices and hundreds of billions of edges on a single commodity machine.
FlashGraph, on a single machine, meets and surpasses the performance of
distributed graph processing engines that run on large clusters.
Furthermore, the small memory footprint
of FlashGraph suggests that it can handle a much larger graph in a single
commodity machine. Therefore, FlashGraph results in a much more economical
solution for processing massive graphs, which makes massive graph analysis
more accessible to users and
provides a practical alternative to large clusters for such graph analysis.

\section{Acknowledgments}
We would like to thank the FAST reviewers, especially our shepherd John Ousterhout,
for their insightful comments and guidance for revising the paper. We also
thank Heng Wang and Vince Lyzinski in Department of Applied Mathematics \& Statistics
of the Johns Hopkins University for discussing the applicatins of FlashGraph.
This work is supported by DARPA N66001-14-1-4028, NSF ACI-1261715, NIH NIBIB 1RO1EB016411-01,
and DARPA XDATA FA8750-12-2-0303.

{\footnotesize \bibliographystyle{acm}
\bibliography{fast15}  % sigproc.bib is the name of the Bibliography in this case

%\theendnotes

\end{document}